\documentclass[final,3p,times,twocolumn]{elsarticle}

\usepackage{hyperref}
\usepackage{verbatim}
\usepackage{url}
\usepackage{amsmath}

\usepackage{multirow}
\usepackage{multicol}

\usepackage{amssymb}

\usepackage{lineno}[displaymath, mathlines]
\usepackage{mathabx}

\usepackage{subcaption}
\usepackage{siunitx}

\journal{Nuclear Instrument and Methods in Physics}

\begin{document}

\begin{frontmatter}



\title{Characterization of Silicon Photomultipliers after proton irradiation up to $10^{12} n_{eq}/mm^2$}


\author[fbk,uniud,tifpa]{Anna Rita Altamura} 
\author[fbk,tifpa]{Fabio Acerbi} 
\author[gsi]{Chiara Nociforo}
\author[fbk]{Veronica Regazzoni}
\author[fbk,tifpa]{Alberto Mazzi}
\author[fbk,tifpa]{Alberto Gola}

\address[fbk]{Fondazione Bruno Kessler (FBK), Sensors \& Devices (SD), via Sommarive 18, I-38123, Trento, Italy}
\address[uniud]{University of Udine, via Palladio 8, 33100 Udine, Italy}
\address[tifpa]{TIFPA - Trento Institute for Fundamental Physics and Applications, via Sommarive 14, I-38123, Trento, Italy}
\address[gsi]{GSI Helmholtzzentrum für Schwerionenforschung Darmstadt - Germany}

\begin{abstract}
Silicon photomultipliers (SiPMs) are highly-sensitive photodetectors emerging as the technology of choice for many applications, including large high-energy physics experiments where they often are exposed to high radiation fluences. In recent years, there has been an increasing interest in assessing the performance deterioration of such detectors after the irradiation with proton or neutron, with different fluence levels.

In this work, samples of different FBK SiPM technologies, made with different manufacturing technologies, were irradiated at the INFN-LNS facility (Italy) with protons reaching fluences up to $10^{12}n_{eq}/mm^2$ (1 MeV neutron equivalent) which correspond $10^{14}n_{eq}/cm^2$ to and their performances were characterized in detail after an approximately 30 days room temperature annealing. 
The results show a significant worsening of the primary noise (dark count rate) of the detectors, which increases with the irradiation dose, whereas the other performance parameters like the micro-cell gain, the correlated noise probability and the photon detection efficiency do not show significant variations over the investigated dose range. The breakdown voltage estimation after irradiation is another important aspect for a SiPM. In this contribution, we show several methods for its estimation and compare the results. We also introduced new methodologies to characterize the performance of the SiPMs when they present a very high level of noise.  

Lastly, we also analyzed the spatial localization of the proton-induced defects inside the device, i.e. the defects that mostly contribute to the increase of the DCR of the device, through the emission microscopy (EMMI) technique. In particular, we analyzed the SiPMs at the single cell level, trying to identify and spatially localize the defects. 
\end{abstract}


\end{frontmatter}


\section{\textit{Introduction}}
\label{S:1}
Silicon Photomultipliers are arrays of many single-photon avalanche diodes (SPADs) connected in parallel to common anode and cathode, each one with an integrated quenching resistor. Each pixel is sensitive to the single photon, thanks to its operation in Geiger-mode, with high internal electric fields triggering self-sustaining avalanche multiplication processes. They are emerging as detector of choice in many applications\cite{Acerbi}, such as nuclear medicine \cite{4436825}, big physics experiments \cite{Garutti_2011}, optical spectroscopy \cite{opticalspectroscopy1,opticalspectroscopy2}, automotive LiDAR \cite{automotive}, etc. 
In particular, when used in high energy physics (HEP) experiments, like for example CALICE Analog Hadron Calorimeter (AHCAL) \cite{calice} and  CMS-HCAL \cite{CMSHCAL,CMSendcap}, or in space applications, SiPMs are often exposed to a significant dose of radiations. Typical values of the maximum radiation doses in such experiments are in the order of $10^{9} - 10^{10} n_{eq}/cm^2$ (1 MeV neutron equivalent, i.e. $10^{7} - 10^{8} n_{eq}/mm^2$) \cite{space} for space applications, or $10^{14} n_{eq}/cm^2$ (i.e. $10^{12} n_{eq}/mm^2$) for calorimetry applications in HEP experiments  \cite{Moll}. 
In order for the SiPMs to work efficiently until the end of the experiment, a good radiation tolerance (or radiation hardness) is required and the SiPMs have to be properly optimized to be able to survive such radiation doses or to minimize the effect of the damage as much as possible, thus reducing their performance worsening due to radiation damage.
In FBK (Trento, Italy) we have been developing different SiPM technologies over the last years, each one optimized for different applications in terms of detection efficiency (for example with peak sensitivity in the blue-wavelength region, or in the green-wavelength region) and performance characteristics in specific application conditions (e.g. cryogenic operations) \cite{LiN}. Having a wide variety of SiPM technology available, it is also very interesting to be able to characterize and compare their main performance-parameters degradation after irradiation. In this way, we can obtain an overview of the radiation hardness of FBK SiPM technologies and we can analyze possible correlations between specific processes / layout splits and their behavior after irradiation.

Generally, as for the majority of silicon-based photodetectors, the radiation damage can affect both the surface and the bulk regions\cite{surface,SiPMRadiationDAmage} through ionizing energy loss (IEL) and not-ionizing energy loss (NIEL) respectively, introducing defects and recombination centers. Depending on the energy and the particle type, either one or the other type of damage is more relevant and likely to happen. As a consequence of the damage the SiPMs show an increase of the not-multiplied current, which we define as the "leakage current" ($I_{leak}$), and the current from the bulk that is multiplied in Geiger mode, which generates the dark counts and is defined as the "dark current" ($I_{dark}$)\cite{Acerbi}. Thus, we observe an increase of the noise and possibly also a decrease of the signal amplitude. In fact, because of the micro-cells "busy" triggering on the dark counts, they are not able to detect events generated by the absorption of photons. This eventually leads to a reduced Signal-to-Noise Ratio (SNR).

In our investigation, protons were chosen for their property of being heavy charged particles doing both ionizing and non-ionizing interactions.

In this contribution, we present and compare the characterization results of several SiPM samples fabricated in FBK with different SiPM technologies (e.g. RGB-HD, NUV-HD) \cite{dcr}. We irradiated the SiPMs with 62MeV protons, in August 2019 at the INFN-LNS in Catania(Italy) at eight fluences steps, from about $1.7 \times 10^{8} $ $n_{eq}/cm^2$ to $1.7 \times10^{14}$ $n_{eq}/cm^2$, corresponding to $10^{6}$ $p/mm^2$ and $10^{12}$ $p/mm^2$, respectively. Devices were irradiated as naked dies, to minimize the effect of the additional irradiation caused by backscattering and secondary effects of the packages and the protective resins. The naked chips are $3\times 3$ $mm^2$ or $4\times 4$ $mm^2$ test structures containing smaller SiPMs, usually $1\times1$ $mm^2$ SiPMs. One chip per each technology type was irradiated to one of the selected proton fluences, to be able to characterize after irradiation different structures with different doses in the FBK labs, in dark and light environments. We characterized the main noise parameters, such as primary noise, gain, crosstalk, etc. In particular, we focused on the bulk defects (which are the main ones with protons at 62MeV). Furthermore, performing some Emission Microscopy (EMMI) tests, we estimated the position of the defects inside the single SPADs and their distribution over the whole SiPM active area.

The irradiation of different technologies could also provide hints on the main features affecting the damage of the SiPMs such as, for example, the internal field, the presence of tranches on the cell borders, etc. 

In this paper, we first describe the experimental setup and some details on the technologies tested. In the subsequent sections, we present the results obtained, focusing, in particular, on the current-voltage (I-V), the detection efficiency, the dark count rate and correlated noise evaluation. Then, the last section describes the emission microscopy analysis.

Since almost all the SiPM tested had a $1\times1 mm^2$ ares, from this point forward we will refer to the fluence in $p/mm^2$ unit, in order to provide the reader with a more immediate idea of the impact of the radiation on the tested devices.

\section{Experimental setup}
\begin{figure}[tb]
\centering
\begin{subfigure}{.47\textwidth}
  \centering
\includegraphics[width=0.8\linewidth]{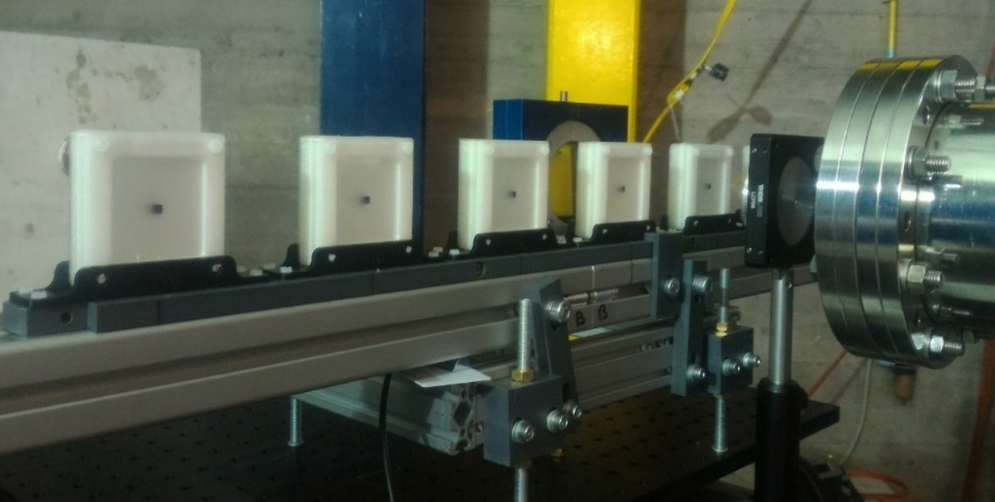}
\end{subfigure}
\begin{subfigure}{.47\textwidth}
  \centering
\includegraphics[width=0.8\linewidth]{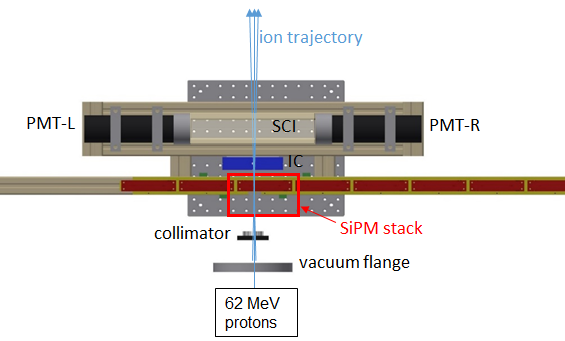}
\end{subfigure}
\caption{Picture and schematic representation of irradiation setup. Along the ion trajectory we placed a collimator, followed by the stack of the SiPM chips, an ionization chamber and two scintillators to measure the effective number of the protons passing through the chips. As showed in the picture, we were able to shift laterally the stack of chip being irradiated to change the one hit by the ion beam.}
\label{fig:setup}
\end{figure}

A schematic representation of the setup is shown in Fig.\ref{fig:setup}. Along the ion trajectory, we placed a collimator (2.7 mm diameter) followed by the stack of SiPM chips, a ionization chamber (IC, 50mm diameter) and a plastic scintillator (SCI, 3 mm thickness) at the end. The output photons were read by blank photo-multiplier tubes (PMT) to measure the effective number of the protons passing through the chips.

\begin{figure}[tb]
\centering
\begin{subfigure}{.23\textwidth}
    \centering
    \includegraphics[width=1\linewidth]{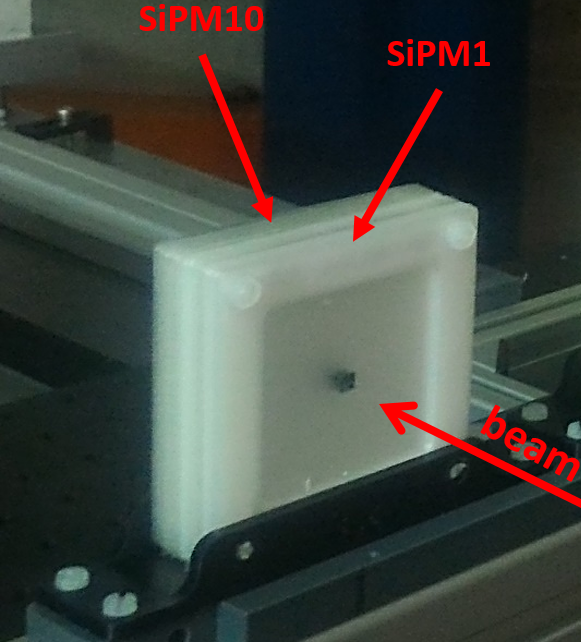}
\end{subfigure}
\begin{subfigure}{.47\textwidth}
    \centering
    \includegraphics[width=7.55cm]{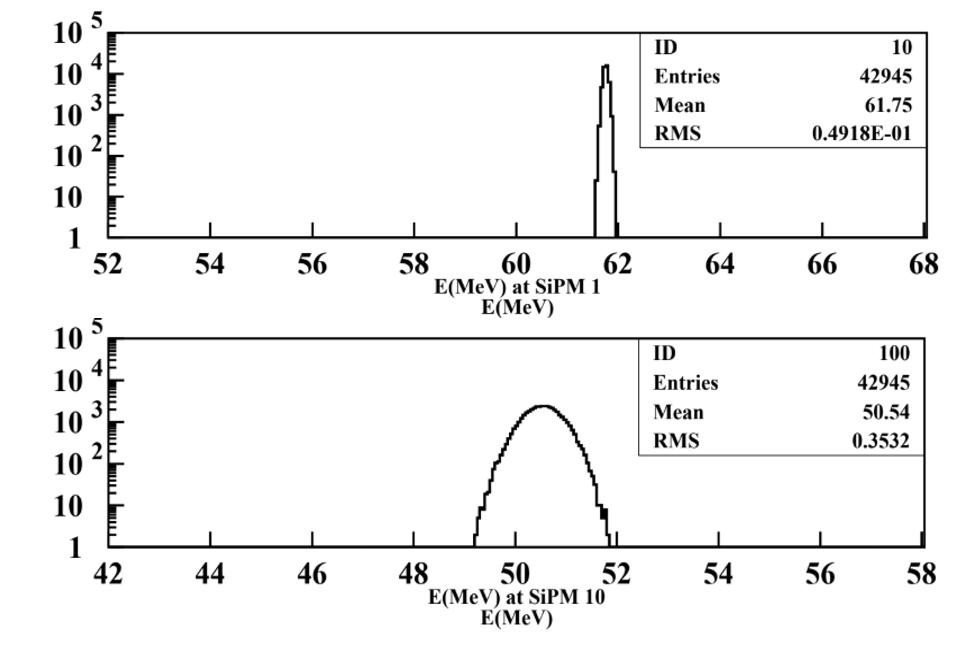}
\end{subfigure}
\caption{Simulated proton energy loss distributions at the first (upper) and the last (lower) SiPM in the stack.}
\label{fig:eloss}
\end{figure}

Different sets of chips from FBK productions were arranged in identical stacks (Fig.\ref{fig:eloss}). One stack per time was inserted in the proton beam line by means of a motorized linear stage and then irradiated. 
Each stack (ten in total) contained 10 layers of SiPM dies, placed one in front of the other. Each chip was placed on a wafer dicing tape inserted in a 3D-printed plastic frame. All the samples in the stack were supposed to receive, simultaneously, the same irradiation dose (see next paragraph).
Eight stacks were irradiated with a different dose, a ninth stack was assembled but left outside the irradiation room, to be used as a reference for estimation of the non-irradiated performance of the SiPMs. Moreover, a tenth stack was left inside the irradiation room throughout the period of irradiation tests but not irradiated, to measure a possible background dose. Indeed, we observed a significant performance variation in the samples. However, it was not possible to estimate the progression of the damage with the fluence but this additional effect should be taken into account to obtain a detailed estimation of the total fluence uncertainty. 

Unfortunately, the manual positioning of the chips inside the frames is subject to a small position uncertainty, estimated in less than $\pm0.5 mm$. This translates into an uncertainty in the effective dose for the SiPMs into the different stacks.

Moreover, this type of stacked-chip setup exposes the beam to a moderate energy loss among the SiPMs into the stack. According to Monte Carlo simulations, the energy difference between the proton energy loss at the first and the last sensors was about 11 MeV, as shown in Fig.\ref{fig:eloss} \cite{mocadi}. The number of protons impinging on each SiPM was estimated thanks to the ionization chamber (IC) and compared with the simulations. The total proton fluence was estimated through the Eq.\ref{eq:num_p} after a calibration of the scaler of the IC. The IC current was integrated over the exposure time and then converted into the number of protons through the IC-SCI calibration and then corrected for a factor 0.97, corresponding to the loss of the protons while passing through the scintillators, according to the Monte Carlo simulations. 
Lastly, the resulting number of protons was multiplied by a different correction factor for each SiPM into the stack which we called "transmission factor $t_{SiPM}$" (see Eq.\ref{eq:num_p}), to take into account the losses caused by the widening of the proton beam along the stack and the alignment of the SiPM active area with respect to the center of the gaussian proton beam.

\begin{equation}\label{eq:num_p}
    N_{ions,SiPM}=\frac{N_{ions,SCI}}{0.97} * t_{SiPM}
\end{equation}

\begin{figure}[tb]
\centering\includegraphics[width=7.5cm]{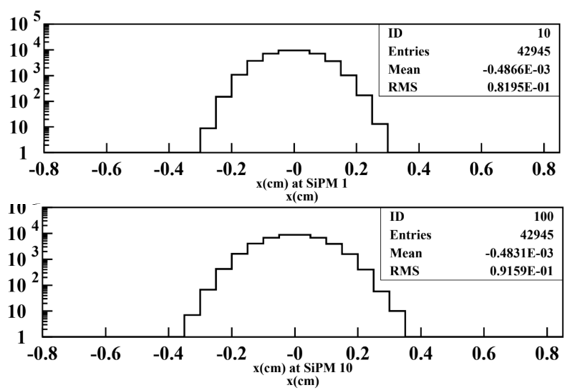}
\caption{Simulated horizontal proton distributions at the first (upper) and the last (lower) SiPM in the stack.}
\label{fig:widthspread}
\end{figure}

\begin{figure}[tb]
\centering\includegraphics[width=6cm]{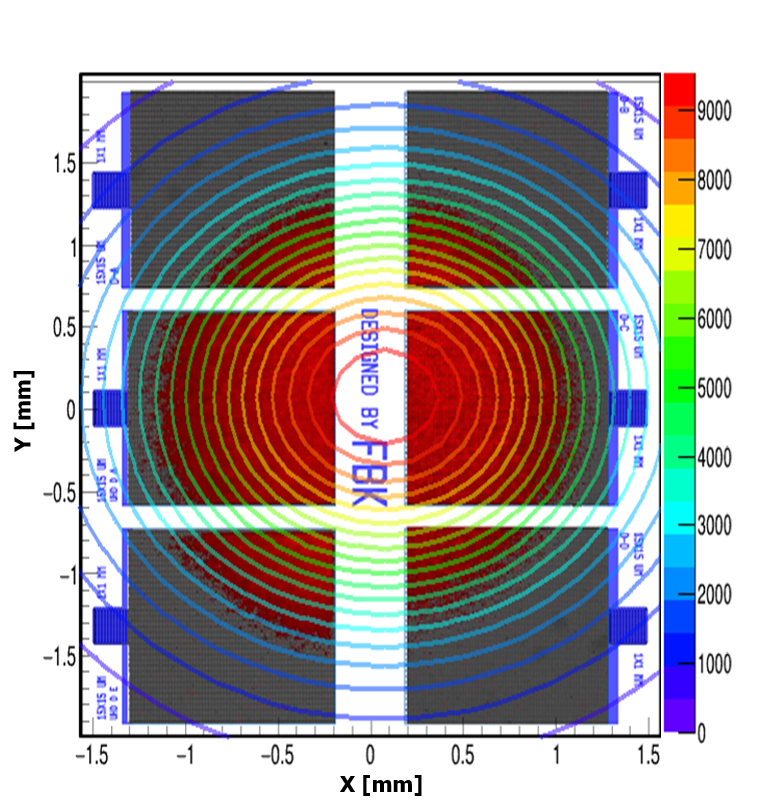}
\caption{Example of superposition of: i) contour lines, representing the estimated irradiation intensity (i.e. beam profile from simulations), ii) emission microscopy (EMMI) intensity, measured with SiPMs biased above the breakdown voltage after irradiation (red color), iii) layout of one irradiated $3\times 4$ $mm^2$ chip, that contains different $1\times 1$ $mm^2$ SiPMs. The EMMI (see Sec.\ref{sec:emmi}) was performed to check the actual center of the Gaussian beam over the chip.}
\label{fig:beamspot}
\end{figure}

Indeed, the beam had a gaussian profile with a 2 mm FWHM in diameter, widening up to 2.5 mm between the first and that the last SiPM in the stack (See Fig.\ref{fig:widthspread}). Assuming the SiPM chips having a variable dimension, between $3\times 4$ $mm^2$ and $4.14\times 4.14$ $mm^2$, the irradiation resulted not uniform over the whole structure. As an example, Fig.\ref{fig:beamspot} shows the simulated beam profile intensity obtained from simulations (contour lines) superposed to the layout of the SiPM chip and to the EMMI image (red colours indicating emission intensity) after irradiation on a single layer of a the stack. As the intensity in the EMMI measurement is proportional to the local DCR and the Gain of the SiPM micro-cells, we used this measure to estimate the actual position of the proton beam during irradiation. Because of the non-uniformity of the beam over the device active area, the final ``effective fluence'' values were estimated separately for each SiPM in each chip of the stack, as an average of the different fluence levels inside the SiPM area. For each chip, the center of the beam was estimated based on the EMMI images of the SiPMs, similarly to what done in the example in Fig.\ref{fig:beamspot}. Then, the effective irradiation dose was calculated as the integral of the beam over the surface of each SiPM. 

After irradiation, the irradiated SiPMs were maintained at room temperature for an annealing time of 30 days. Then, several functional measurement were performed.

\subsection{Tested Silicon Photomultipliers}
Several SiPMs were irradiated, each one having a different layout and made with a different technology. Despite all of them were measured in terms of reverse current-voltage curves, functional characterizations were performed only on a subset of them, to check the main variations in their functional parameters. We compared the NUV-HD technology (peak sensitivity in the near ultraviolet, high density of cell, with p-on-n junction type, on n-type of substrate) \cite{gola}, RGB-HD technology (green-peaked sensitivity, high density of cells, with n-on-p junction type and p-type substrate)\cite{dcr}, NUV-HD-LF (near ultraviolet sensitive with n-type substrate, but with lower electric field, with higher breakdown voltage) and RGB-UHD-LF (p-type substrate, with ultra high density of cells and very small cell pitch \cite{AcerbiUHD2018}, with lower field and higher breakdown voltage), as summarized in Table\ref{tab:tech}.

\begin{table}[h]
\centering
\begin{tabular}{|l|l|l|l|l|l|l|}
\hline
Technology & cell pitch & FF & $V_{bd}$ & $PDE_{pk}$ 
\\
 & [$\mu$m] & [\%] &  [V] &  [nm] 
\\
\hline
RGB-HD & 20 &  66 & 28.5  & 535 
\\
\hline
RGB-HD & 25 & 72  & 28.5  & 535 
\\
\hline
RGB-HD & 30 & 78  &  28.5 & 535 
\\
\hline
NUV-HD & 30 &  78 & 28.8  &  420    
\\
\hline
NUV-HD & 35 & 80  & 28.8  & 420 
\\
\hline
NUV-HD & 40 & 81  & 28.8  &  420   
\\
\hline
NUV-HD-RH & 15 & 51  & 32.2  & 400 
\\
\hline
RGB-UHD LF & 10  & 68  & 34.4  & 535
\\
\hline
RGB-UHD LF & 12.5  &  74 & 34.4  & 535
\\
\hline
\end{tabular}
\caption{Summary of the key features of the tested SiPMs technologies, at $+20^{\circ}C$. Reference for RGB-HD in \cite{Acerbi_RGB-HD}, NUV-HD in \cite{Piemonte2016,gola}, RGB-UHD in \cite{AcerbiUHD2018}. }
\label{tab:tech}
\end{table}

\begin{figure*}[tb]
	\centering
		\begin{subfigure}{.33\textwidth}
		\includegraphics[width=\textwidth]{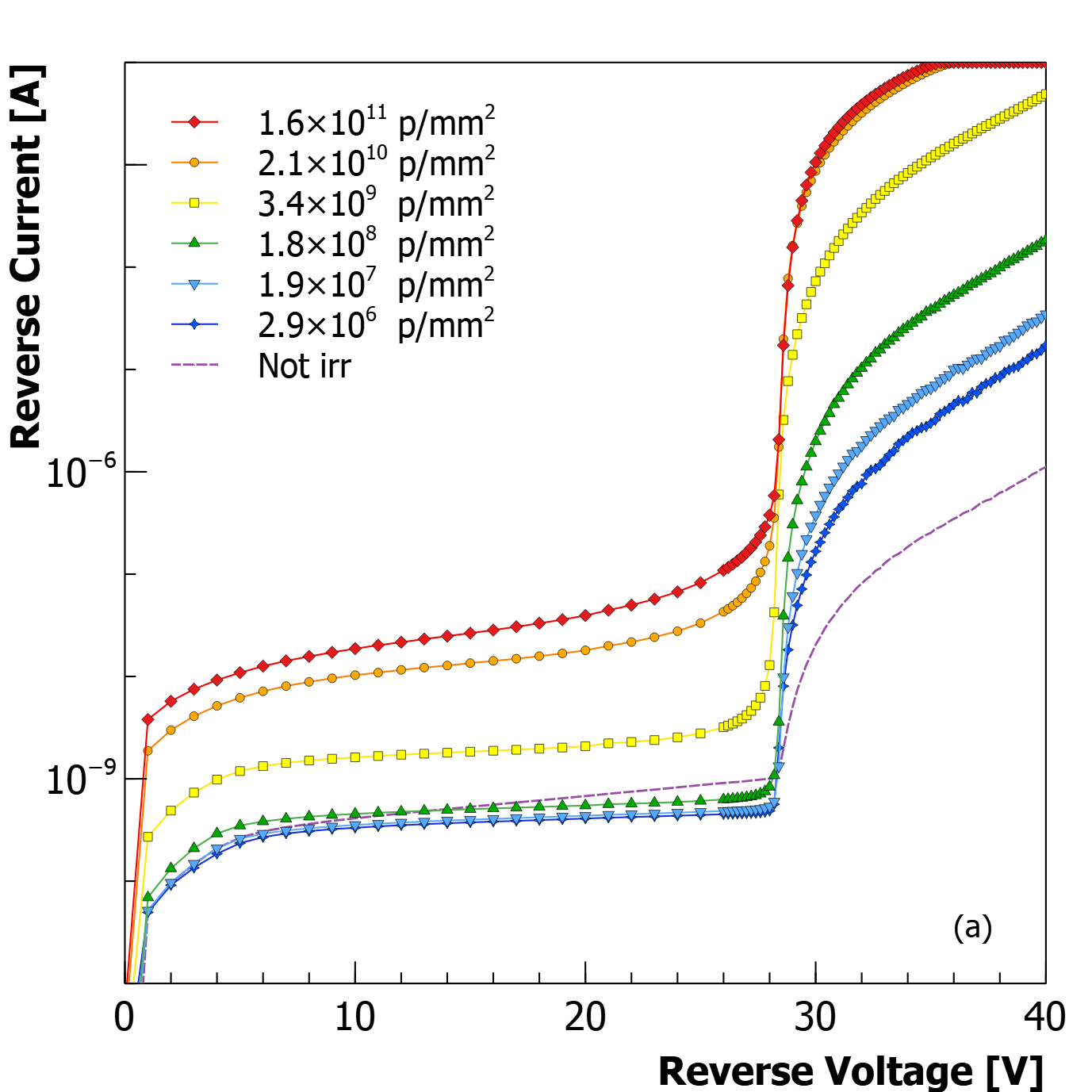}
	\end{subfigure}
	\begin{subfigure}{.33\textwidth}
		\includegraphics[width=\textwidth]{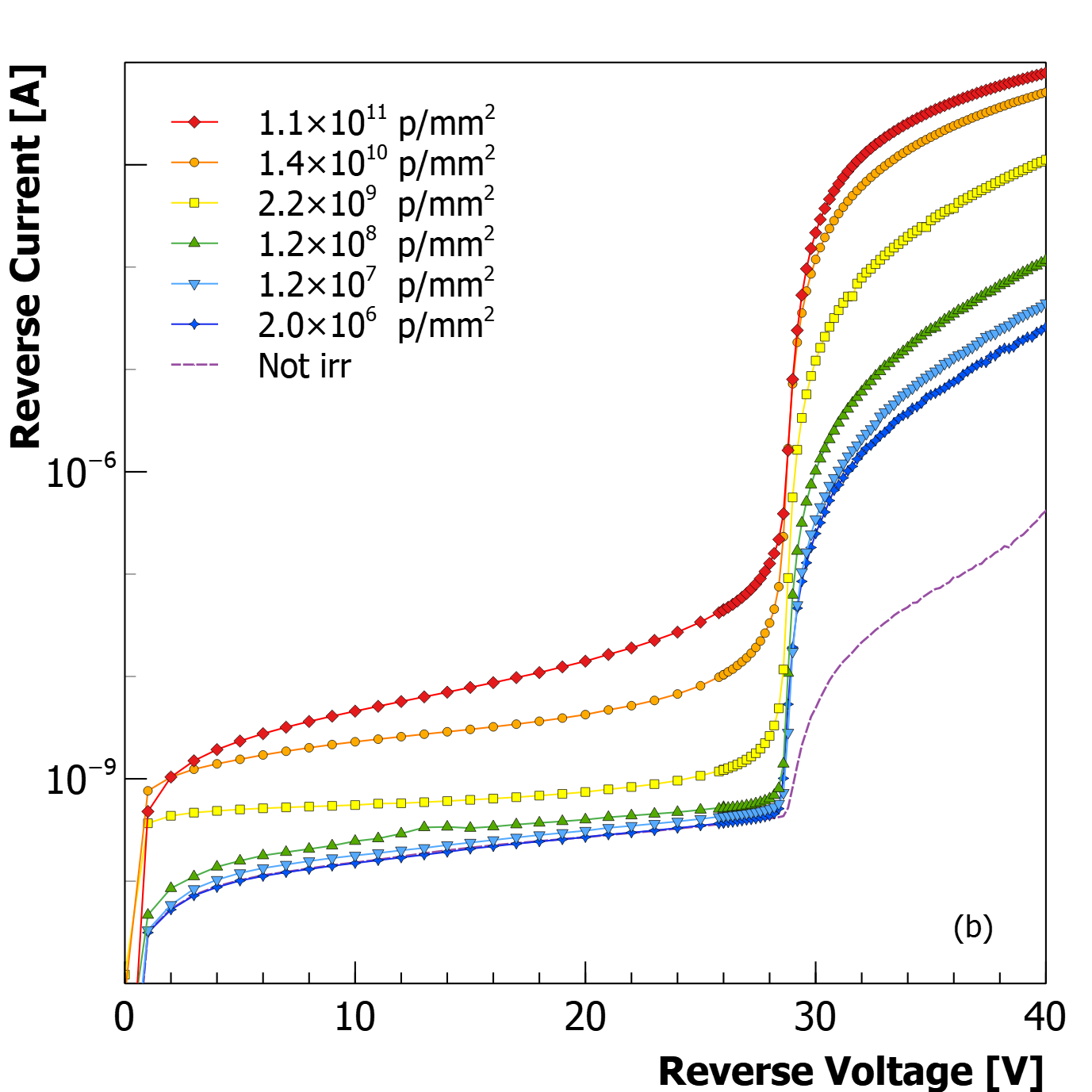}
	\end{subfigure}
	\begin{subfigure}{.33\textwidth}
		\includegraphics[width=\textwidth]{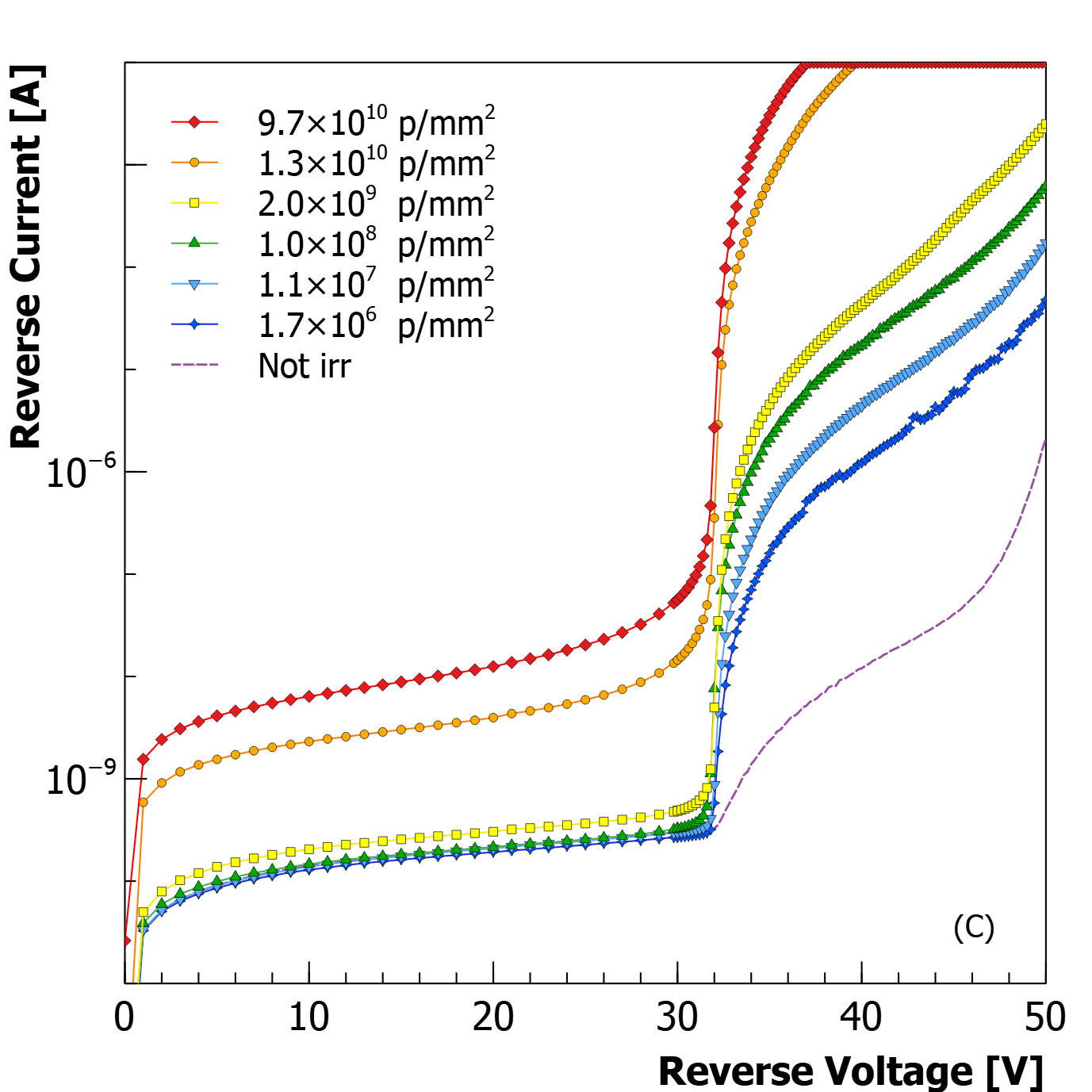}
	\end{subfigure}
	\caption{Current-voltage curve of NUV-HD SiPM with 35$\mu$m cell pitch (a), of RGB-HD SiPM with 25 $\mu$m cell pitch (b) and of RGB-UHD-LF SiPM with 10 $\mu$m cell pitch, at different irradiation fluences.}
	\label{fig:compNUVandRGB}
\end{figure*}

\begin{figure}[tb]
	\centering
		\includegraphics[width=0.47\textwidth]{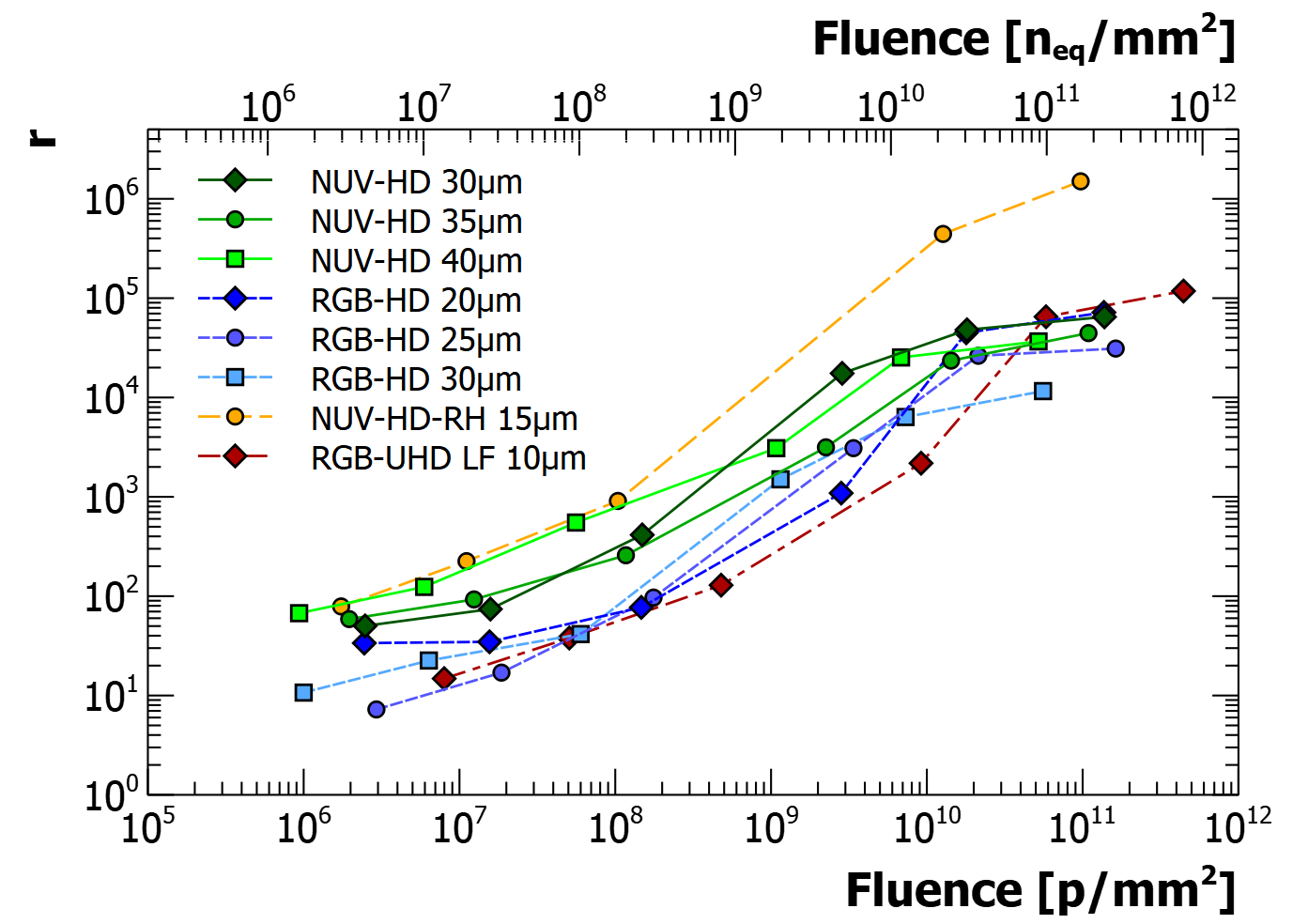}
	\caption{current ratio (r) trend as a function of the fluence at 5V excess bias}
	\label{fig:r}
\end{figure}

\begin{figure*}[tb]
	\centering
	\begin{subfigure}{.48\textwidth}
		\includegraphics[width=1\textwidth]{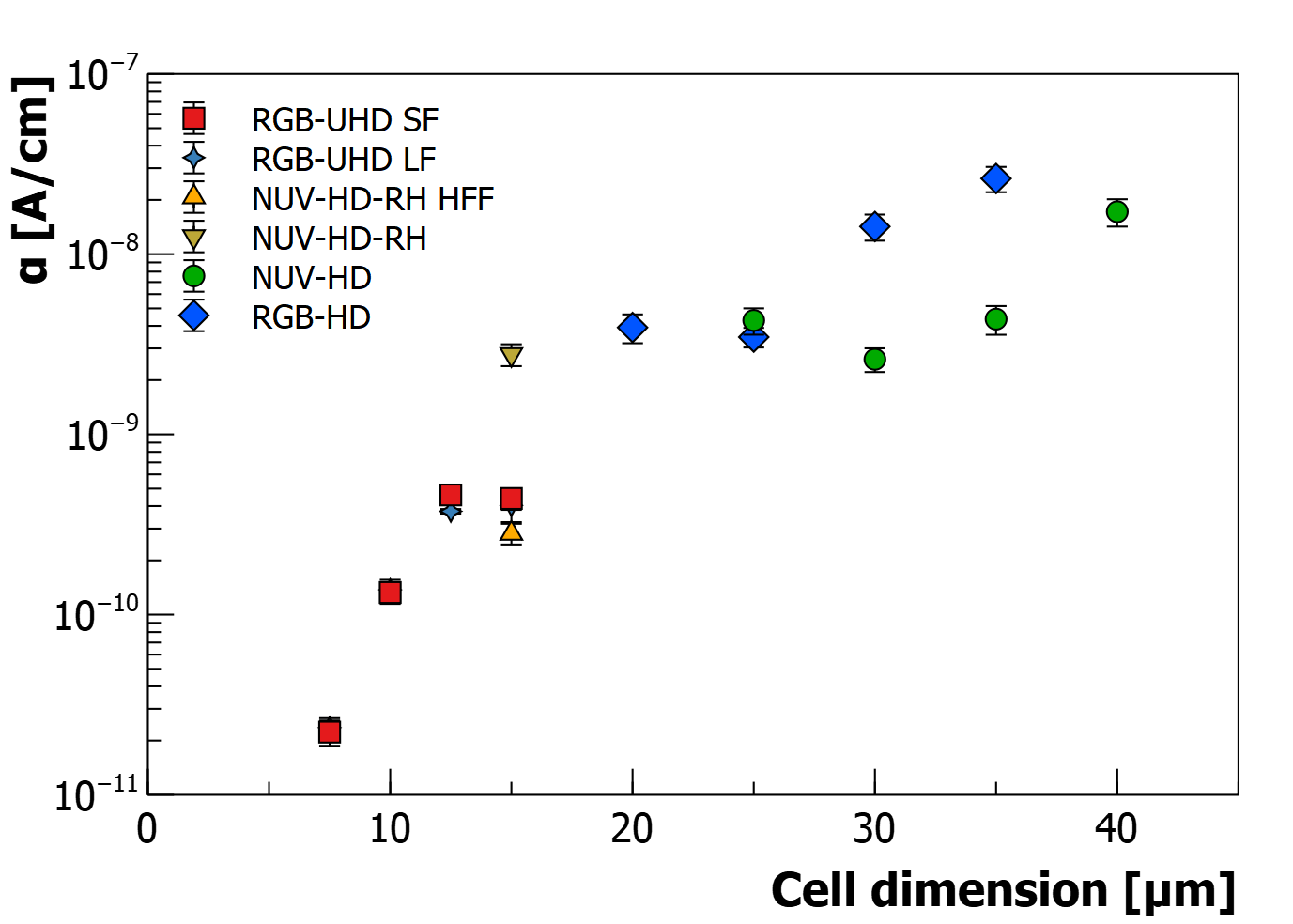}
	\end{subfigure}
	\begin{subfigure}{.48\textwidth}
		\includegraphics[width=1\textwidth]{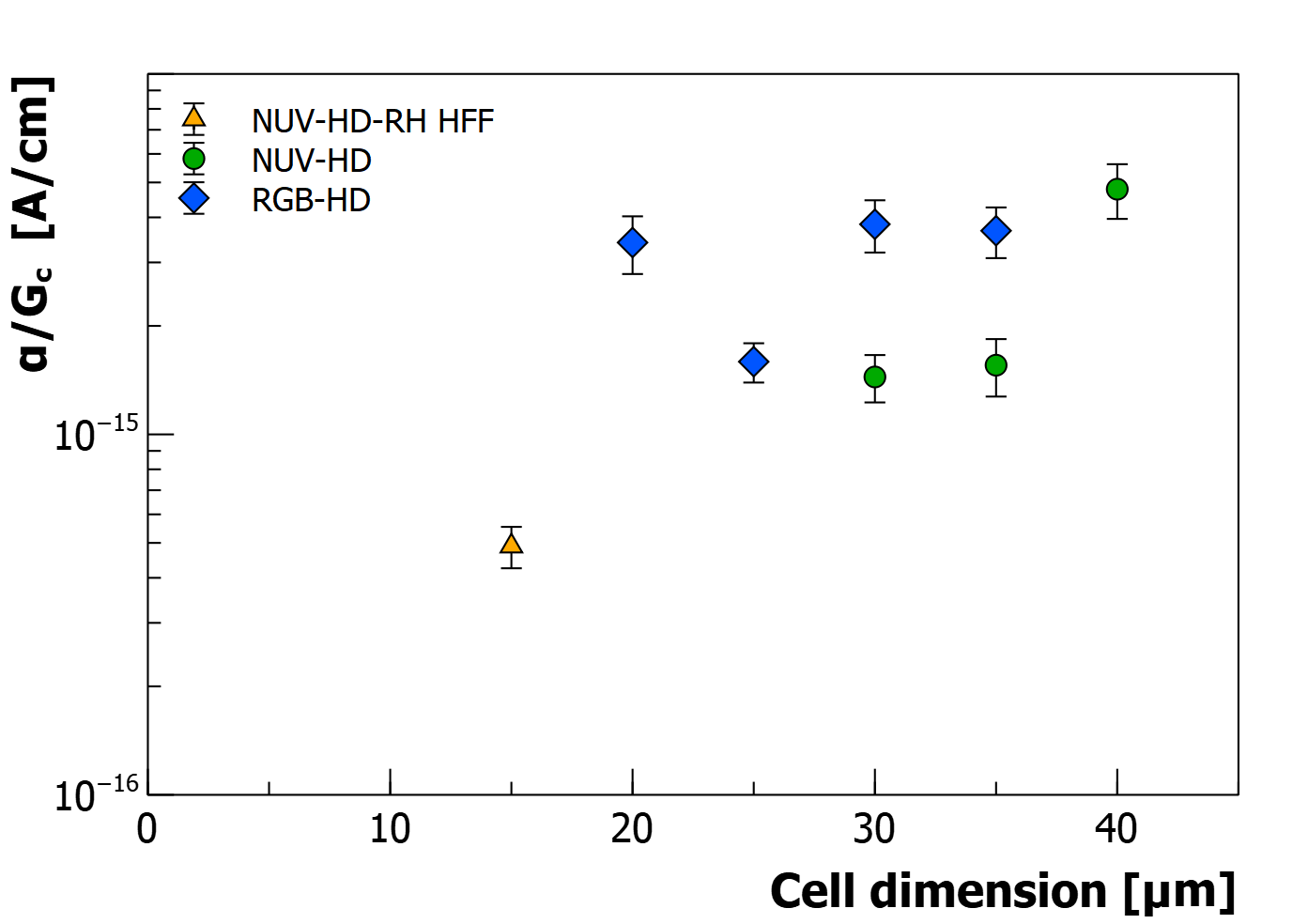}
	\end{subfigure}
	\caption{Damage parameter $\alpha$ plotted as a function of the cell dimension of the SiPMs, for SiPMs at +$20^{\circ}C$, biased at 5 V overvoltage (i.e. over breakdown voltage) without any normalization (left) and normalized to the micro-cell gain (right).}
	\label{fig:damage}
\end{figure*}

\section{Results}
Before introducing in this section the main results of the irradiation tests, a brief discussion on the approach to the measurement methods needs to be done. In fact, the radiation damage in the silicon sensors with an internal amplification has not theoretical basis as solid as in the case of silicon sensors without an internal gain.

The measurement methods in this paper were inspired by the typical approaches used for the silicon sensors without any internal amplification. In most cases, these were adjusted or replaced by mew methods to take into account the gain and the avalanche triggering probability, which is the probability for a carrier to start an avalanche.

\subsection{Reverse current-voltage curves}\label{sec:IV}
One of the key measurement for the study of the performances of a SiPM is the current in reverse bias which includes two main contributions: the ``leakage current'' $I_{leak}$ (the pre-breakdown not-multiplied current) and the ``dark current'' $I_{dark}$ (the post-breakdown current, multiplied by avalanche multiplication, due to the spurious avalanche triggering).

Fig.\ref{fig:compNUVandRGB} shows the I-V curves after 30-day annealing at room temperature of NUV-HD, RGB-HD and RGB-UHD SiPMs, with 35 $\mu$m, 25 $\mu$m and 10 $\mu$m cell pitch respectively. 
All the samples showed an increase of both $I_{leak}$ and $I_{dark}$ as expected \cite{SiPMRadiationDAmage}. 
To better quantify the worsening of the performance we introduced a new parameter "\textbf{r}" defined as the ratio between the current after irradiation (at a specific fluence) and before irradiation, thus representing the relative increase of the current after the specific irradiation:
\begin{equation}
    r=\frac{I_{after}(V_{ex})}{I_{before}(V_{ex})}
\end{equation}

where $V_{ex}$ is the excess bias, i.e. the difference between the bias and the breakdown voltage.

As a first approximation, the current ratio \textbf{r} provides an estimation of the noise increment due to irradiation, as we will see later in the text, and it gives information about the evolution of the damage with the fluence.

In Fig.\ref{fig:r}, the current ratio \textbf{r} is plotted as a function of the fluence, at 5 V above the breakdown voltage. A linear trend between $10^{9}$ $p/mm^2$ and $10^{11}$ $p/mm^2$ can be observed, whereas the trend is different at higher fluences. There seems to be first an approximately linear behaviour, followed by a saturation effect starting at about $10^{10}$ $p/mm^2$. From the figure we can see an increment between more than four orders of magnitude and six orders of magnitude in post-breakdown bias region.

An alternative way to estimate the damage in the bulk, which is typically used also for radiation detectors without gain, is the "current-related damage factor $\alpha$"\cite{DamageMoll:1999kv}:
\begin{equation}
\Delta I = \alpha \Phi V
\end{equation}
where $\Delta I$ is the difference between the reverse current before and after the irradiation, $\phi$ is the fluence and V is the volume of the depletion region within the detector. This equation correlates the current with the fluence, without saturation or second order effects, thus assuming the defects generation into the bulk as not-correlated to each other, the value of $\alpha$ is supposed to be constant at the different fluences, assuming there are no charge multiplication mechanisms. In the case of sensors with an internal amplification like the SiPMs, the parameter $\alpha$ cannot be directly compared with the one of the sensors without an internal gain. This is beacuse beacuse the current include the internal gain of the micro-cells and also because there are field enhancement effects in the deep levels, created by the proton irradiation. Thus, as a first approach, we have at least divide $\alpha$ by the micro-cell gain. Here we used the so-called "Current Gain" ($G_c$), which will be explained in detail later in the text. Fig.\ref{fig:damage} shows a comparison between the results of the damage parameter estimated with and without the micro-cell gain normalization. We can see that $\alpha$ increases with the cell dimension in the first plot, but this is due to the different gain. When they are normalized to the micro-cell gain as in the second plot, we do not observe a particular trend. 

\subsection{Breakdown voltage estimation}
The breakdown voltage ($V_{bd}$) is defined as the reverse voltage where the multiplication factor, which is the number of couples of carriers generated through impact ionization after a single initial couple, diverges. Thus it represents the ``ideal'' starting point of the Geiger-mode operation (see \cite{Acerbi}), for detectors like SPADs and SiPMs. Many parameters of the SiPM (e.g. gain, DCR, PDE) depend on the excess bias, i.e. the reverse bias in excess to the breakdown voltage. Therefore, $V_{bd}$ is a key parameter and it is important to asses possible variations with respect to the irradiation. 
However, its estimation in SiPMs is not always straightforward. 
For an accurate estimation of the $V_{bd}$ from the reverse current-voltage curve, we need a clear distinction between the pre and post breakdown parts. In this sense, its estimated value can be altered when $I_{leak}$ is comparable to the value of $I_{dark}$ at low excess bias, either because $I_{leak}$ is very high or because $I_{dark}$ is particularly low. In these cases, it can  be useful to extract the breakdown voltage from the reverse current-voltage curves obtained under a faint illumination. Too high light intensities should be avoided because the increasing current in the linear multiplication regime (i.e. the one of the avalanche photodiode, APD, few volts below breakdown voltage) could distort the curve, resulting in a worsening of the discrimination between pre and post breakdown curve and leading to a premature and wrong breakdown estimation. 

In literature several methods have been used by different authors for the $V_{bd}$ estimation \cite{Vbd}, \cite{Acerbi}, \cite{Garutti_2011}.
In our experimental characterization, we compared several of them based on the reverse I-V characteristic curve of the SiPM or the pulse-shape analysis. They were compared to study their accuracy as a function of the irradiation dose. The methods that were considered are: (i) the maximum of the First Logarithmic Derivative (FLD), (ii) the maximum of the Second Logarithmic Derivative (SLD), (iii) the minimum of the Inverse of the Logarithmic Derivative  (ILD), (iv) the maximum of the Normalized First Derivative (NFD), (v) the bias at pulse-amplitude equal to zero (A), (vi) the bias at Gain equal to zero (considering only the first two points of the gain vs bias curve) ($G_2$), (vii) the bias at Gain equal to zero (considering the whole curve) ($G_{all}$). 

In the first two methods the $V_{bd}$ is estimated from the voltage corresponding to the maximum of the first or the second derivative of the Log(I) vs V curves, i.e. qualitatively the maximum slope point or the inflection point of the reverse current-voltage curve, with a logarithmic vertical axis. Both methods rely on the fact that the gain of the SiPM micro-cells increases rapidly when the reverse bias exceeds the breakdown voltage, switching from a gain of a few thousands, in the linear multiplication regime, to a gain of a few hundreds thousand, in the Geiger regime, in the span of a few hundred of millivolts. Accordingly, the reverse current rapidly increases and the breakdown voltage can be estimated with one of the two methods mentioned above. We also note that, because the Gain in Geiger mode is proportional to the cell capacitance, thus to its area, the difference between the Gain in linear and in Geiger regime is more pronounced for larger SiPM micro-cell sizes. These methods might be less accurate for smaller cell sizes or, as mentioned above, when $I_{leak}$ is particularly high, compared to $I_{dark}$. 
The ILD method \cite{KLANNERVbd} considers the minimum of the inverse of the first logarithmic derivative. The NFD method was used in \cite{MUSIENKOVbd} and it is based on the derivative of the ``normalized current-voltage curve'', which is the first derivative of the reverse current-voltage curve, normalized to the current value.

On the other hand, the pulse-amplitude, the $G_2$ and the $G_{all}$ approaches are based on the measurements in pulse counting mode. In particular, with the pulse-amplitude method, we consider the peak amplitude of the SiPM single cell pulses, i.e. the signals generated when a single dark count triggers an avalanche, at different biases. Then, the breakdown voltage is calculated as the bias at which the linear extrapolation of the data intercepts the horizontal axis.  With the $G_{all}$ method, we employ a similar extrapolation based on the measured SiPM Gain at different bias, while with the $G_2$ method the linear regression considers only two values of the Gain measured at low excess bias. $G_{all}$ is a common method used in some experiments to calibrate the detectors. However, as described in \cite{Acerbi}, in some devices the gain vs bias curve might not be linear, thus it can be useful to take into account only the first two points of the curve, to avoid the part where the slope changes.
Tipically, with non-irradiated SiPMs we use the SLD, the pulse-amplitude method and the methods based on gain-vs-bias and we normally find them to be in good agreement. However, this might not be the case for irradiated samples. 

In our analysis, we considered the reverse I-V characteristics curve obtained in both dark conditions and under moderate illumination. Indeed, considering the increment in leakage current because of the radiation damage, it might be possible to see a fictitious modification of the breakdown voltage, leading to a worse distinction between pre and post breakdown current. This might shift the estimation with SLD to lower biases, but this is not supposed to happen with moderate light because of the significant increase of the current above the breakdown too. However, depending on the light intensity, SLD might be affected by the above mentioned increment of current in linear-multiplication regime.

\begin{figure}[tb]
	\includegraphics[width=0.47\textwidth]{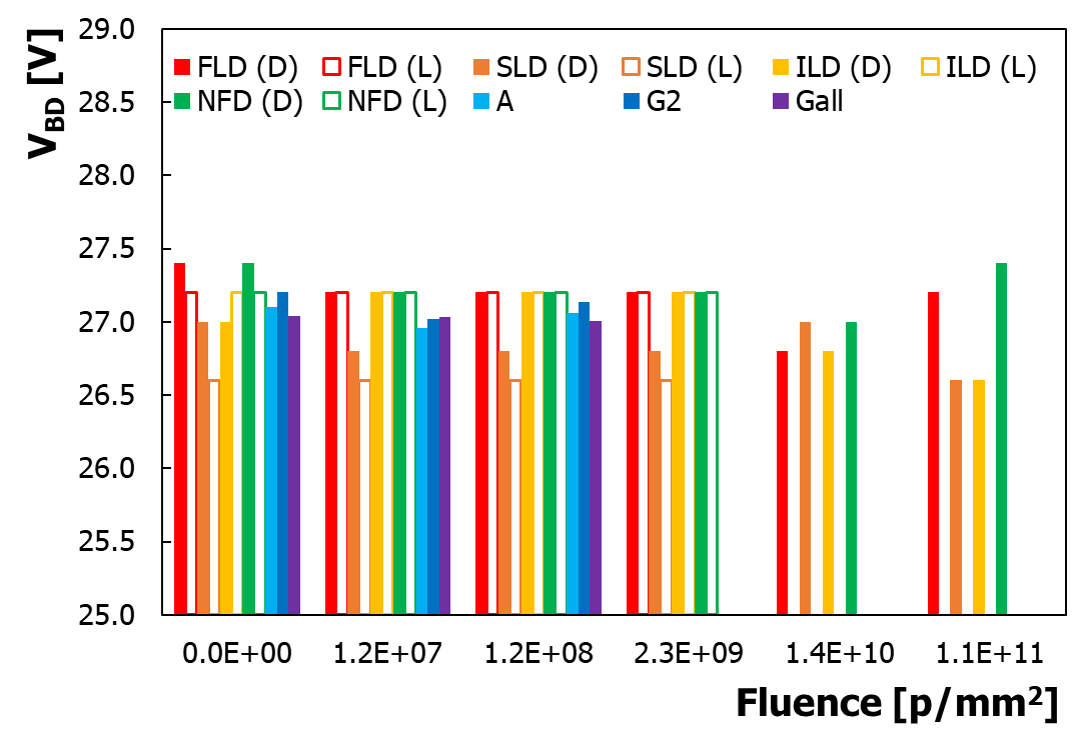}
	\caption{Bar chart of the $V_{bd}$ values for the NUV-HD SiPM with 35$\mu$m cell pitch, obtained with the different estimation methods in dark (D) and light (L) conditions at -$20^{\circ}C$.}
	\label{fig:Vdbnuv}
\end{figure}

A graphic comparison of the different methods for the NUV-HD technology is provided in Fig.\ref{fig:Vdbnuv}, where the light and dark breakdown voltage estimations are shown as a function of the irradiation fluence at -$20^{\circ}C$.
In dark conditions, we can notice a relatively high spread and a discrepancy between the FLD and ILD methods. On the contrary, when using moderate light, ($420nm$ LED) we found smaller variation values. Since the reverse I-V curves do not rise as steep as in the ideal case, the voltage at the second derivative of each curve differs from the one at the first derivative.
Due to the difficulty in estimating the amplitude and gain of the irradiated samples (as described in detail in the next section), the $V_{bd}$ values obtained with these approaches have some fluctuations, especially in the estimation from the gain. Specifically, in the NUV-HD SiPM with 35$\mu$m cell pitch, the pulsed-light method remains efficient only up to $2.3 \times 10^{9}$ $p/mm^2$.

Overall, we can see no trends in breakdown voltage as a function of fluence in all the approaches considered, with some fluctuations mostly in dark conditions.
Thus, we can consider the junction and the doping concentration as not affected by the irradiation up to $10^{11}-10^{12}$ $p/mm^2$.

\subsection{Photon Detection Efficiency} \label{sec:pde}

\begin{figure}[tb]
	\centering
	\begin{subfigure}{.45\textwidth}
		\includegraphics[width=1\textwidth]{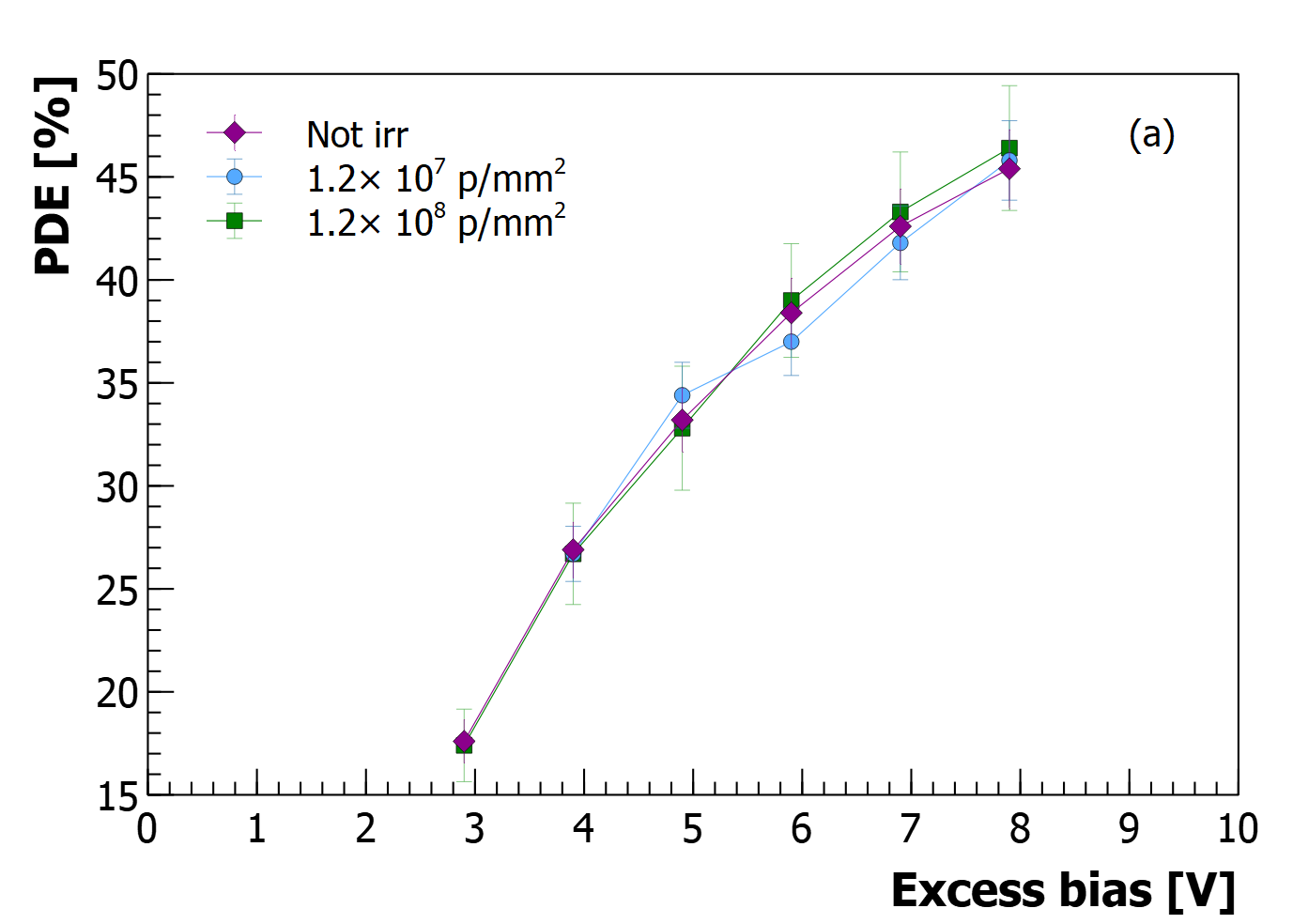}
	\end{subfigure}
	\begin{subfigure}{.45\textwidth}
		\includegraphics[width=1\textwidth]{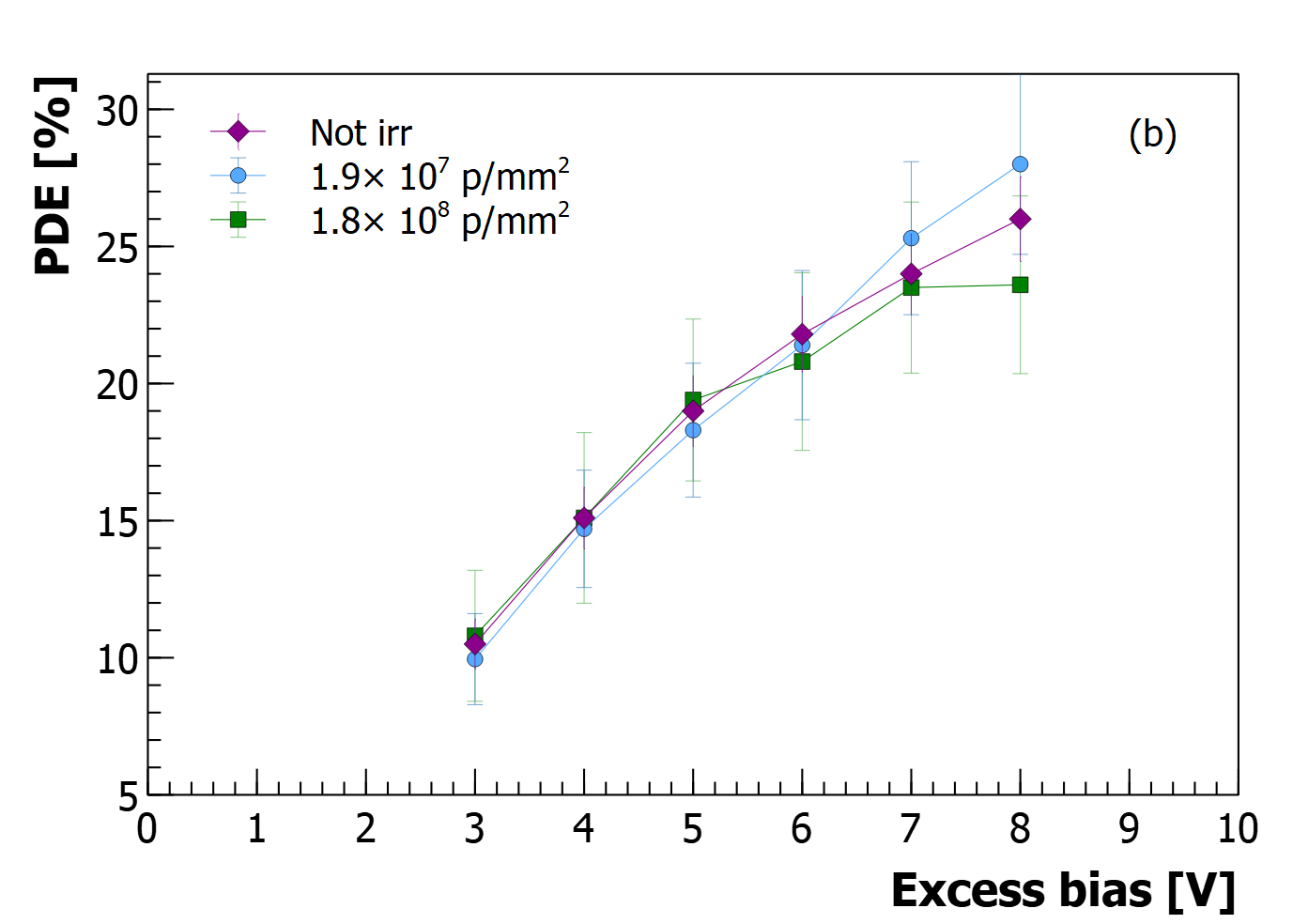}
	\end{subfigure}
	\caption{PDE plots for the NUV-HD SiPM with 35 $\mu$m cell pitch (a) and the RGB-HD SiPM with 25 $\mu$m cell pitch (b), measured on non-irradiated and two irradiated devices under a 420nm LED light, at +$20^{\circ}C$.}
	\label{fig:pde}
\end{figure}

\begin{figure}[tb]
	\centering
	\begin{subfigure}{.45\textwidth}
		\includegraphics[width=1\textwidth]{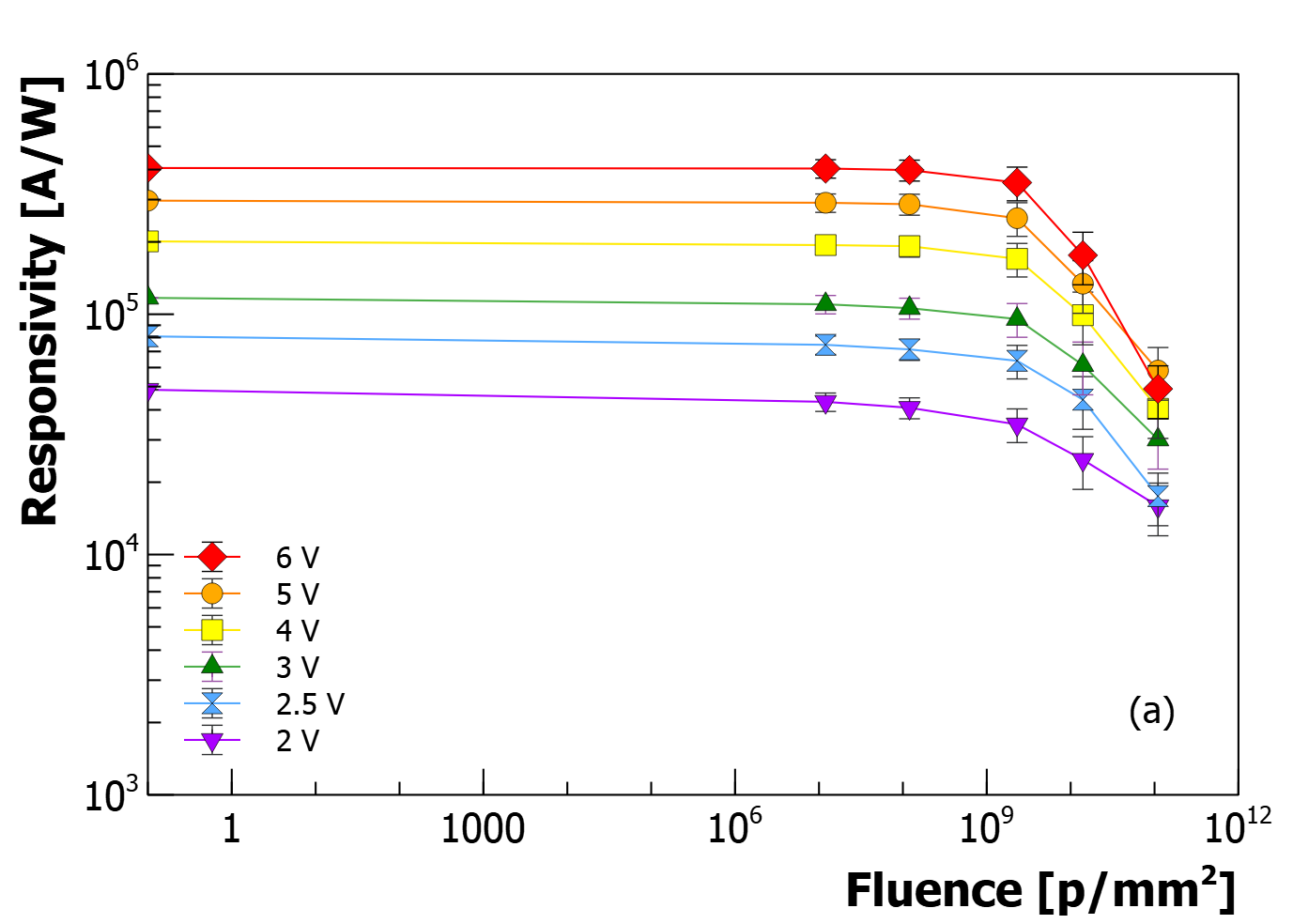}
	\end{subfigure}
	\begin{subfigure}{.45\textwidth}
		\includegraphics[width=1\textwidth]{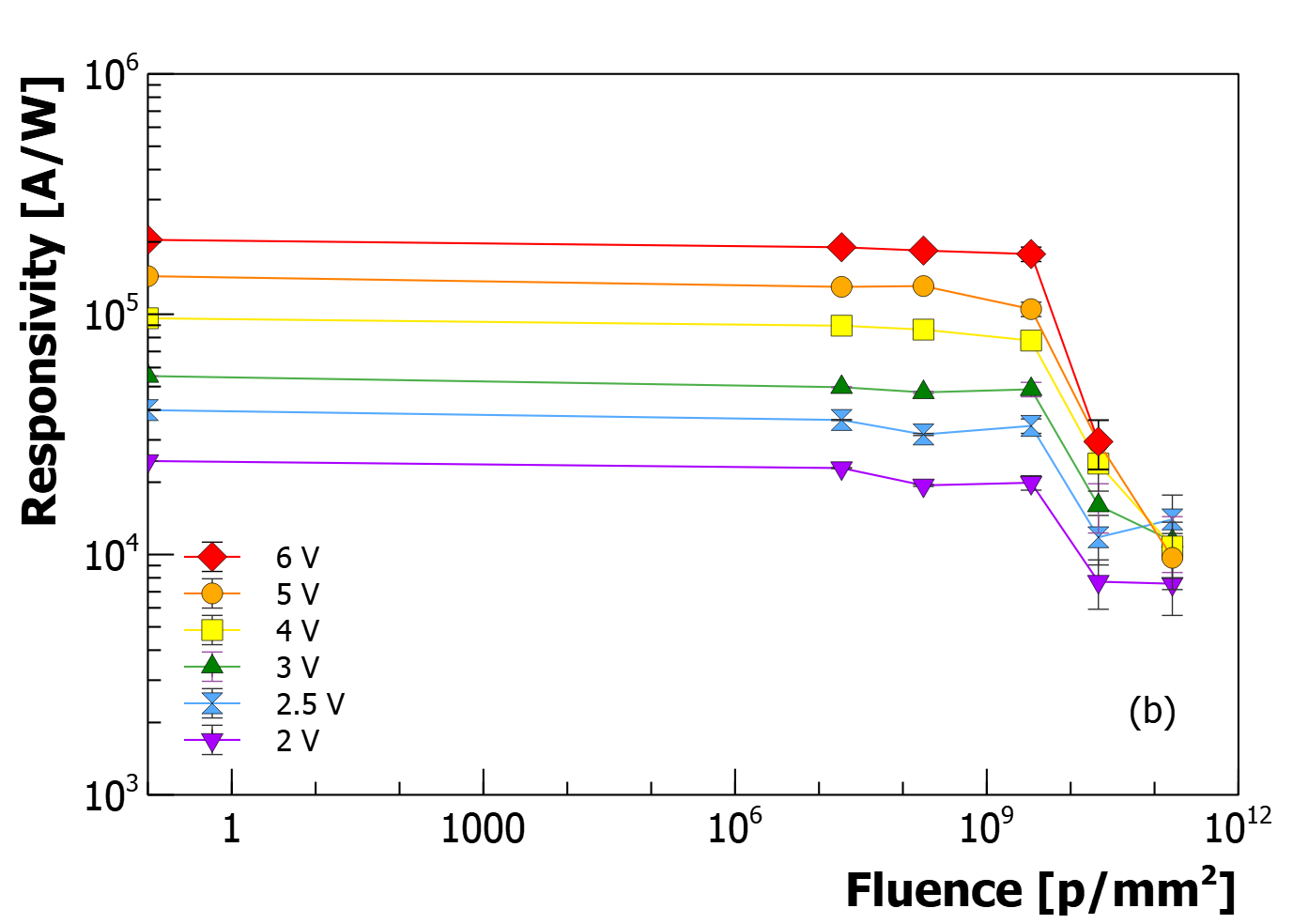}
	\end{subfigure}
	\caption{Plots of the Responsivity as a function of the protons fluence, for the NUV-HD SiPM with 35 $\mu$m cell pitch (a) and the RGB-HD SiPM with 25 $\mu$m cell pitch (b) at $+20^{\circ}C$ and at several excess bias values.}
	\label{fig:responsivity}
\end{figure}

The Photon Detection Efficiency (PDE) is the probability of a photon to be detected by a Geiger-mode detector like the SiPMs. It is the product of the geometrical fill-factor (FF) of the micro-cell, the quantum efficiency (QE), i.e. the probability of the photon to be absorbed in the useful depleted region of the micro-cells and $P_{trig}$, i.e. the probability of the photo-generated carriers to start an avalanche \cite{Zappal2016}:
\begin{equation}
PDE = FF \times QE \times P_{trig}
\end{equation}

Because FF is a layout parameter, we can safely assume that it does not change with the irradiation, whereas some investigations need to be done on the other parameters. 

For the measurement of the PDE, we used the pulsed-light method described in \cite{Zappal2016}. This is based on counting of the number of avalanche pulses within a certain time window, with and without synchronous, calibrated, pulsed light illumination. 

According to \cite{Zappal2016}, the PDE value is calculated as:
\begin{equation}
    PDE=\frac{N_{L}-N_{D}}{(I_{L}-I_{D})\times L_{cal}}
\end{equation}

where $N_{D}$ is the number of events, generated in the dark, within the time window, because of the thermal generation and tunneling,  whereas $N_{L}$  is the number of events generated in light in the same time window, thus is composed by the sum of noise and photo-generation events. $I_{L}$ and $I_{D}$ are the current of the reference diode in light and dark conditions respectively and $ L_{cal}$ is a calibration factor dependent on the geometry of the setup and the area of the SiPM under test. 

We experimentally observed that this method resulted to be usable up to $10^{8}p/mm^{2} $, whereas at higher fluences the noise becomes too high, introducing too many pulses in the evaluation window (which is typically tens of nanoseconds wide) so that the variance of the poissonian stochastic variable $N_D$ is comparable to $N_L$. In this condition, it is difficult to have a good accuracy of the numerator in the PDE formula.

In Fig.\ref{fig:pde} the PDE curves for NUV-HD and RGB-HD SiPMs are plotted for irradiation doses up to $10^{8} p/mm^{2}$. The figure does not show any evident change with respect to the non-irradiated samples. The worsening at high voltages, especially in the RGB-HD plot, is due to the increase of the DCR, affecting the measurement accuracy. In general, we did not expect significant changes in the PDE at high fluences, except in case of a very high DCR, i.e. when the cell occupancy increases considerably.

Another method to evaluate possible variations in the PDE is the measurement of the SiPM \textit{Responsivity}, done in current mode with the some setup used for the PDE measurement, but measuring the output current and not the pulses, Results are shown in Fig.\ref{fig:responsivity} for the RGB-HD and the NUV-HD SiPMs. We can see clearly that up to $10^{10}$ $p/mm^2$ we have no variations, whereas at higher fluence the cell saturation (cell occupancy), reduces the responsivity, thus also the PDE.





\subsection{Current Gain} \label{subs:gain}
To study possible variations of the Gain of the SiPMs, we used a setup based on a light emitting diode (LED), operating in pulsed mode. The light emitted from a 420nm LED was injected into an optical fiber, illuminating the SiPM under test, placed inside a thermostatic chamber, settled at $-20^{\circ}C$. The SiPM was connected to a custom transimpedance amplifier and the amplified signal was read out by a digitizing oscilloscope (Keysight, 10 GSa/s, 1GHz bandwidth). We measured the signals from many LED pulses and averaged them. After measuring the average signal, we also recorded an average noise signal, i.e. the baseline, and then we subtracted it to the first one. The total charge (expressed in electrons) was derived from the formula:
\begin{equation}
    Q =\frac{1}{q} \frac{INT_{net}}{G_{ampli}}
\end{equation}
where $INT_{net}$ is the integral of the net signal (dark signal subtracted) and $G_{ampli}$ is the trans-impedance gain of the amplifier corresponding to 5000 V/A.
The number of photons was estimated first using the not-irradiated SiPM, by taking into account the detection efficiency and the Current Gain:
\begin{equation}
n_{ph}=\frac{Q}{PDE \times G_{c}}
\label{eq:nphotons}
\end{equation}
The "Current Gain"  is the gain is the gain of the micro-cell of the SiPM multiplied by the excess charge factor (ECF). The current flowing through the device in dark conditions can be simply obtained by multiplying the DCR by the "Current Gain" and the electron charge. 

Then, after estimating $N_{ph}$, we measured the average signal of the irradiated SiPMs.

From eq.\ref{eq:nphotons}, we estimated the Current Gain of different SiPMs, at different irradiation doses, assuming that the number of photons emitted by the LED and the PDE were constant. We have verified that the PDE does not change with irradiation dose up to $10^{10}$ $p/mm^2$.

\begin{figure}[tb]
	\centering
	\begin{subfigure}{.47\textwidth}
		\includegraphics[width=1\textwidth]{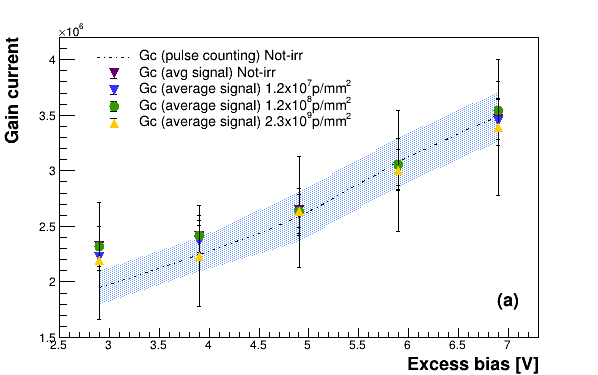}
		\label{ }
	\end{subfigure}
	\begin{subfigure}{.47\textwidth}
		\includegraphics[width=1\textwidth]{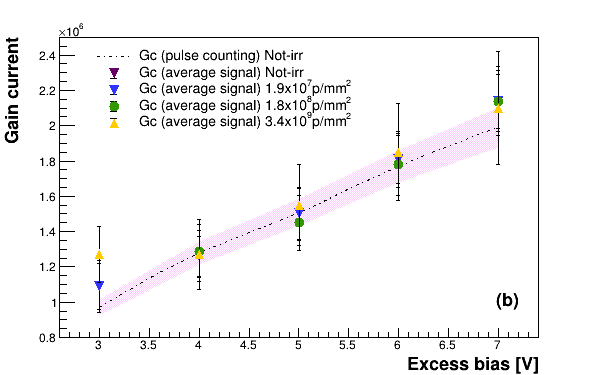}
	\end{subfigure}
	\caption{Current Gain measured directly with pulse counting method, compared to the Current Gain estimated by the average-signal method, for different SiPMs with different irradiation fluences, for the NUV-HD SiPM with 35$\mu$m cell pitch (a) and RGB-HD SiPM with 25$\mu$m cell pitch (b). The error bands refers to the one measured directly with pulse counting method.}
\label{fig:gaincurrent}
\end{figure}

Fig.\ref{fig:gaincurrent} shows the results of the Current Gain measured with the pulse counting method on the not-irradiated SiPM, compared to the Current Gain estimated by the average-signal method, described above, up to $10^9$ $p/mm^2$. It can be seen that the results are in agreement inside the uncertainty range, indicating that overall the PDE and the Current Gain values of the SiPMs do not change significantly with irradiation, in the investigated range.

\subsection{Dark Count Rate}

As previously indicated, the estimation of the DCR for irradiated SiPM, with high noise levels, can be complicated because of the difficulty in identifying the pulses, thus extracting the amplitude and inter-arrival times which are necessary when using the approach described in \cite{dcr}. 
Thus, we estimated the DCR with two methods: the former is based on the dark current (D.C. method), quantified by the r parameter (from section \ref{sec:IV}), the latter is based on the pulse-identification analysis (P.I. method), using the inter-arrival times between events and their amplitude, as described in \cite{dcr}.

\begin{figure}[tb]
	\centering
	\begin{subfigure}{.45\textwidth}
		\includegraphics[width=1\textwidth]{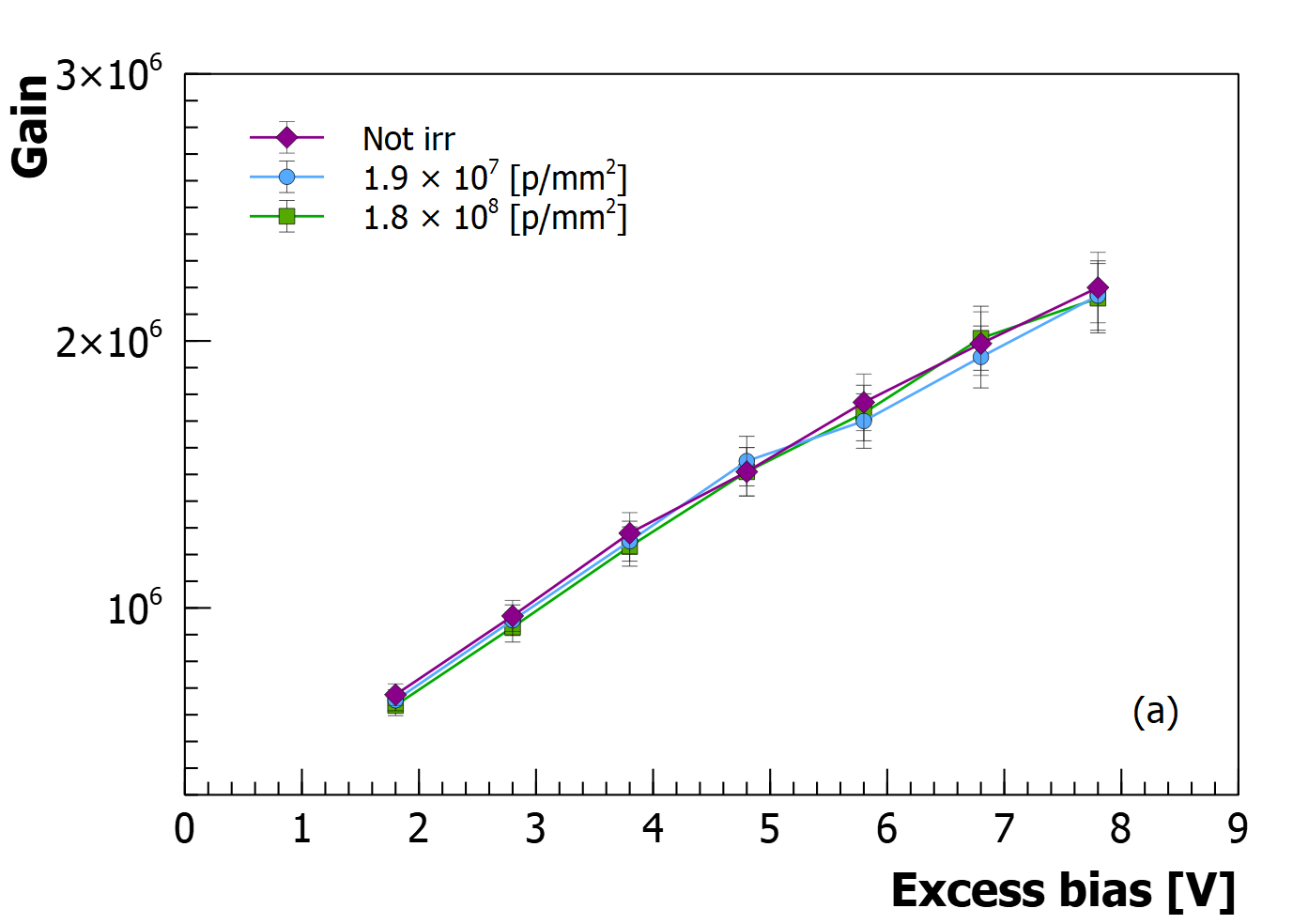}
	\end{subfigure}
	\begin{subfigure}{.45\textwidth}
		\includegraphics[width=1\textwidth]{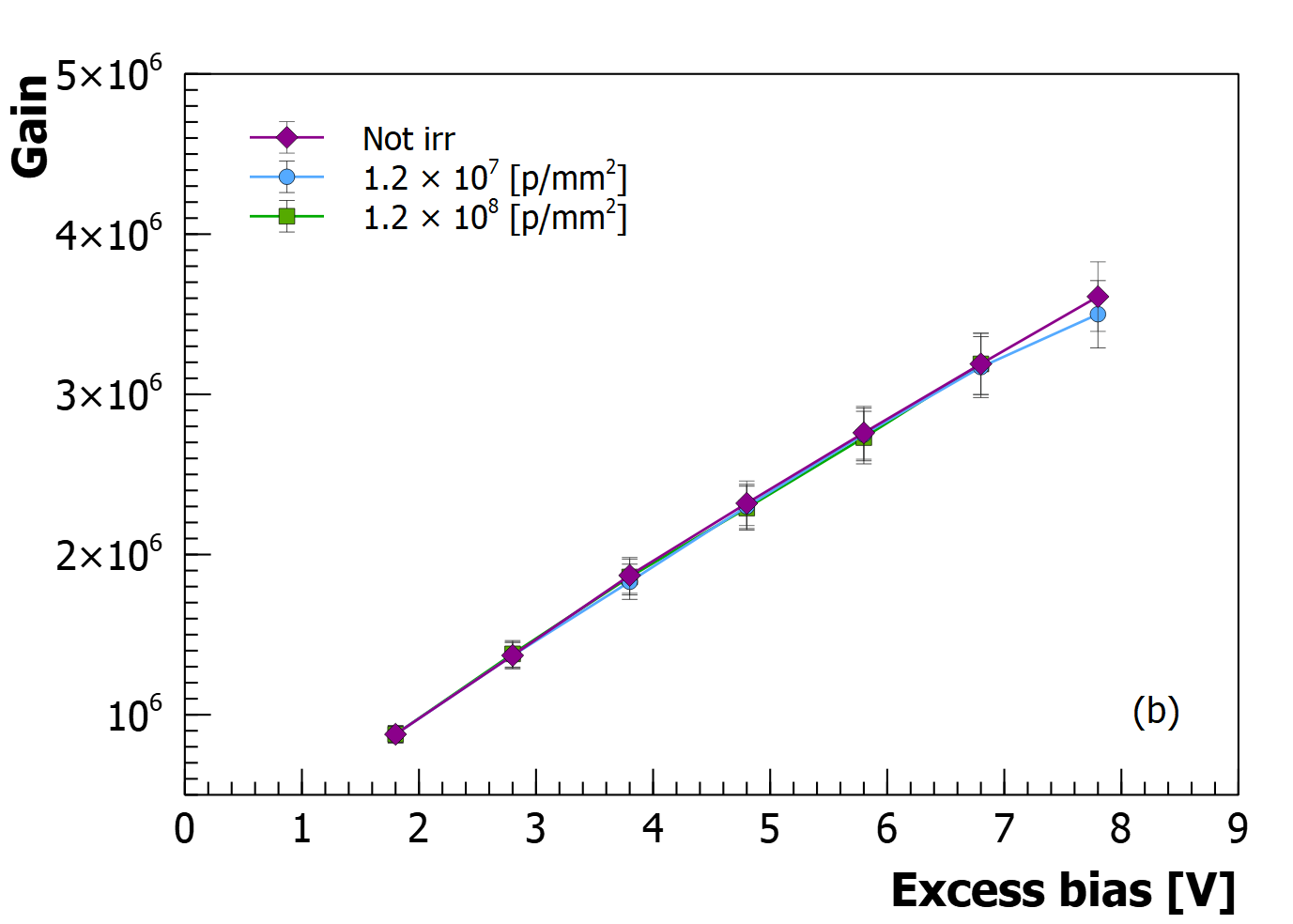}
	\end{subfigure}
	\caption{Gain of the RGB-HD SiPM with 25$\mu$m cell pitch (a) and NUV-HD SiPM with 35$\mu$m cell pitch (b) at two fluence levels.}
	\label{fig:Gain}
\end{figure}

As for the first method, the DCR can be calculated as:
\small
\begin{equation}
\hspace{-0.5cm}
DCR(V_{ex}) = \frac{I_{dark}(V_{ex})}{q\times G(V_{ex}) \times ECF(V_{ex})} = \frac{1}{q} \frac{I_{dark}(V_{ex})}{G_{c}(V_{ex})}
\label{eq.noise.1}
\end{equation}
\normalsize

where $q$ is the elementary electron charge, $G$ is the gain of the micro-cell, the ECF represents the excess charge factor and $G_c$ is the Current Gain. As shown by Eq.\ref{eq.noise.1}, $G_c$ can be used to estimate the primary, Poisson-distributed DCR from the reverse I-V measured on the SiPMs.
In our first method (i.e. equation \ref{eq.noise.1}) we could not calculate separately the DCR and the ECF, thus we supposed the ECF, as well as the Current Gain, are constant (see previous paragraph). This hypothesis is partially confirmed by the measure of the Gain of the SiPMS with the pulse-counting method, which remains constant for all the SiPM tested up to $10^{8} p/mm^2$ at least, as showed in Fig,\ref{fig:Gain} for the RGB-HD and the NUV-HD SiPMs. 

Therefore, assuming $G_{c}$ not changing with fluence, we get:
\begin{equation}
    \frac{DCR(V)_{after}}{DCR(V)_{before}} = \frac{I_{after}}{I_{before}} = r
\end{equation}

This method resulted to be very useful in high fluence conditions, where the noise may increase too high for an accurate distinction between signal and noise. 

The pulse-identification method, instead, is based on inter-arrival times and amplitudes of each avalanche pulse. With this method, we can separate primary events from correlated noise.

\begin{figure}[tb]
	\centering
	\begin{subfigure}{.45\textwidth}
		\includegraphics[width=1\textwidth]{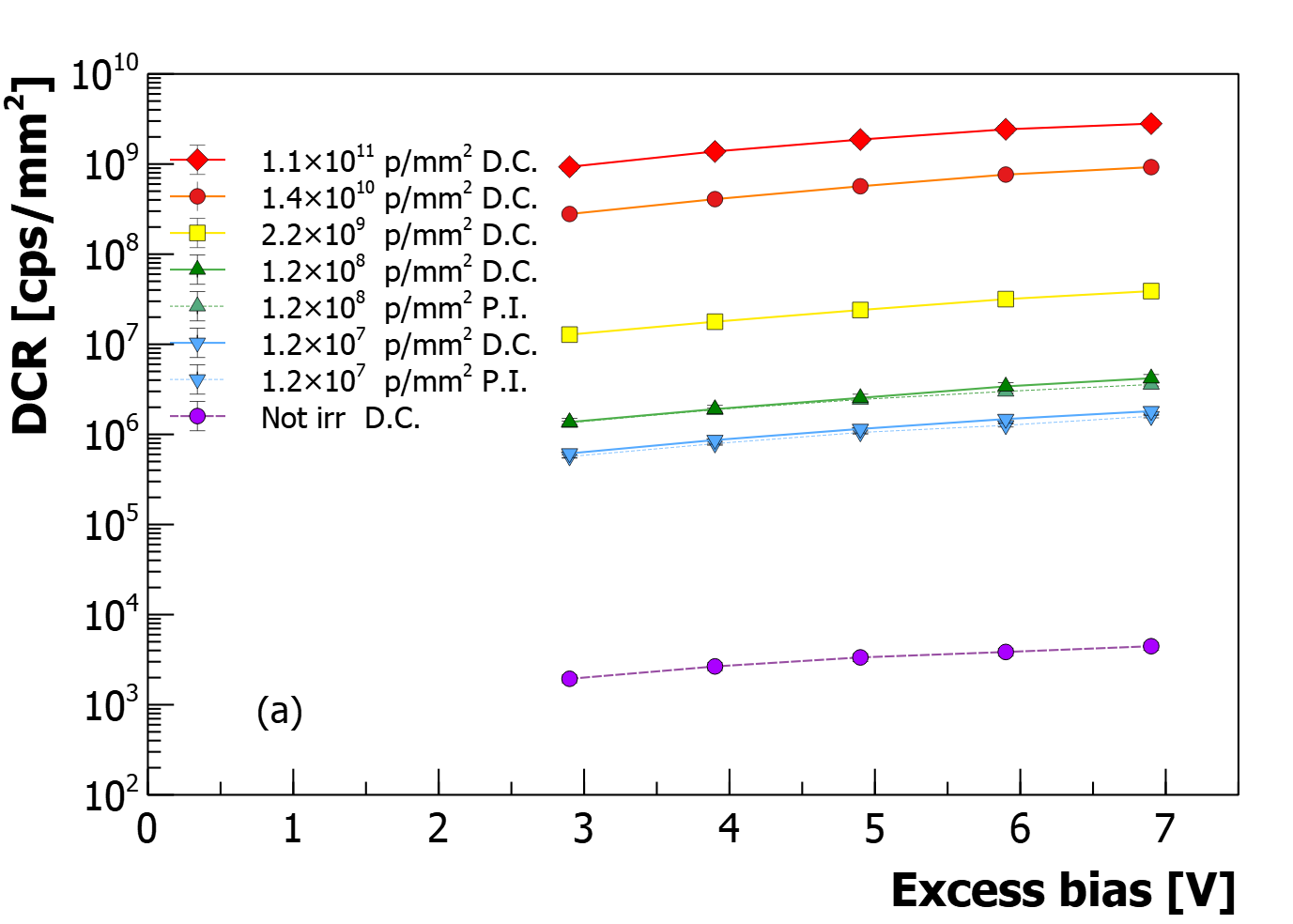}
	\end{subfigure}
	\begin{subfigure}{.45\textwidth}
		\includegraphics[width=1\textwidth]{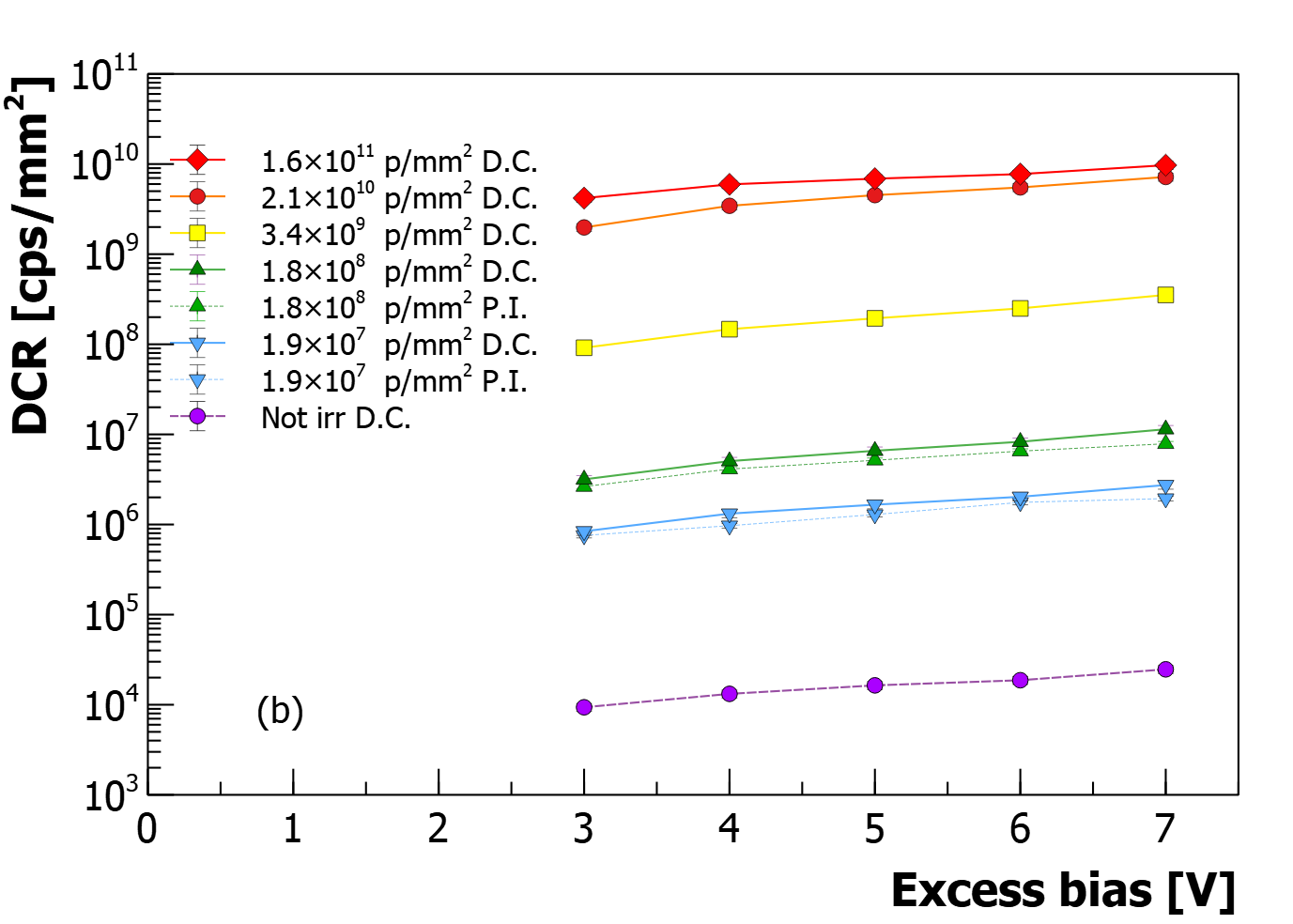}
	\end{subfigure}
	\caption{DCR as a function of the applied voltage estimated with the two methods (i.e. D.C. and P.I. methods) for the NUV-HD SiPM with 35$\mu$m cell pitch (a) and the RGB-HD SiPM with 25$\mu$m cell pitch (b).}
	\label{fig:DCR1}
\end{figure}

In Fig.\ref{fig:DCR1} we show a comparison of the DCR measurement for the NUV-HD technology obtained with both the described methods. Considering the uncertainty range, a good agreement between the results with the two methods can be noticed, at least up to $10^8$ $p/mm^2$, which represents the limit for the pulse counting method in this specific technology.

\begin{figure}[tb]
	\centering
		\includegraphics[width=0.47\textwidth]{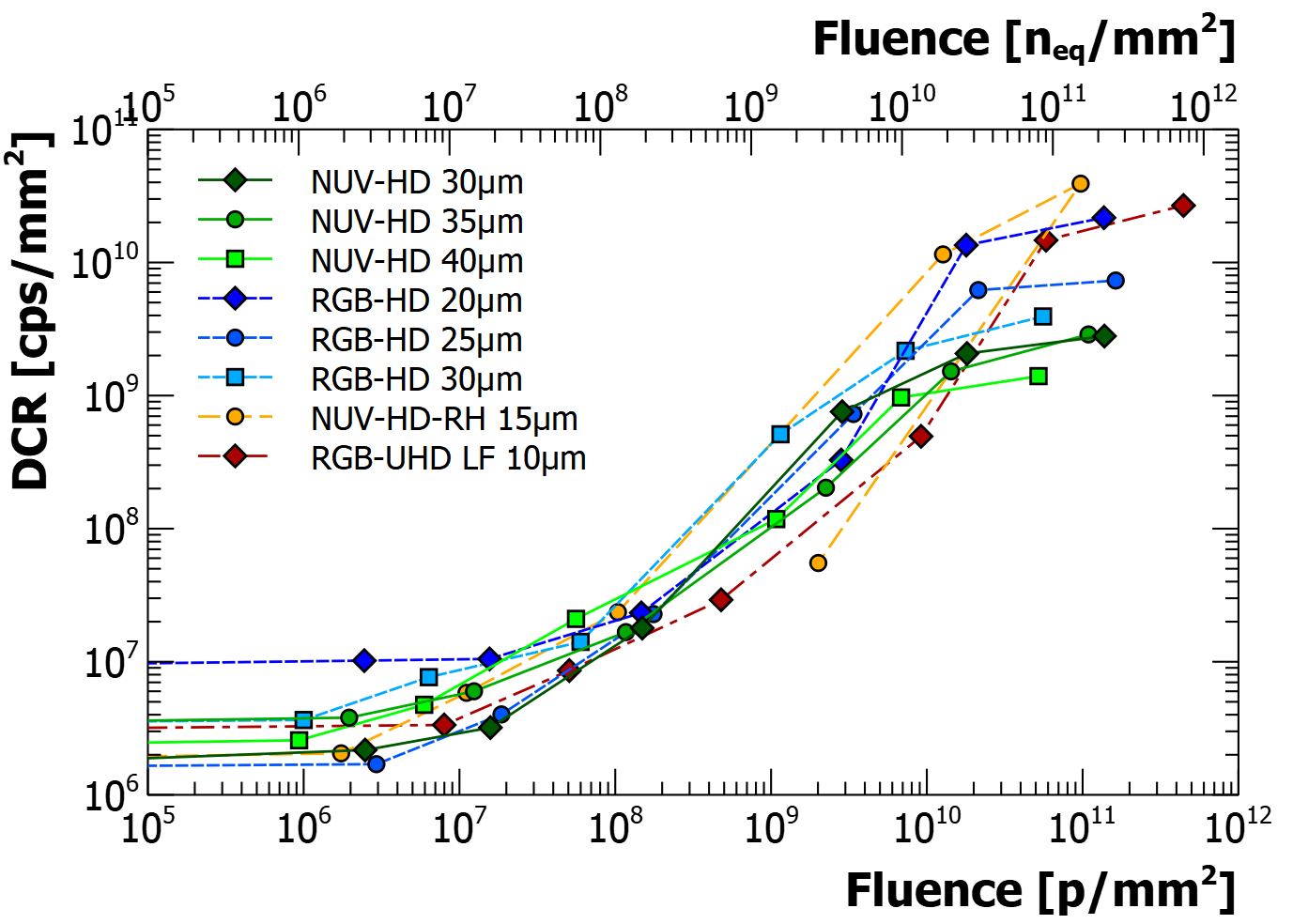}
	\caption{Estimated primary dark count rate (DCR) as function of the fluence at 5V of excess bias}
	\label{fig:DCR_from_r}
\end{figure}

In Fig.\ref{fig:DCR_from_r} the results of the DCR estimation from the dark current approach is shown. The trend of the curves reflects what already seen for the current ratio \textbf{r} in Fig.\ref{fig:r} with a visible saturation effect at high fluences. The plot DCR vs fluence provides an accurate comparison between the technologies. Indeed, the technologies that have an high DCR when not-irradiated, show a smaller DCR increment with the radiation in the low fluence range, being the proton damage partially covered by the native deep-level concentration, which gives the higher DCR. 

\subsection{Correlated noise}\label{subs:correlatednoise}
\begin{figure}[tb]
	\centering
	\begin{subfigure}{.45\textwidth}
		\includegraphics[width=1\textwidth]{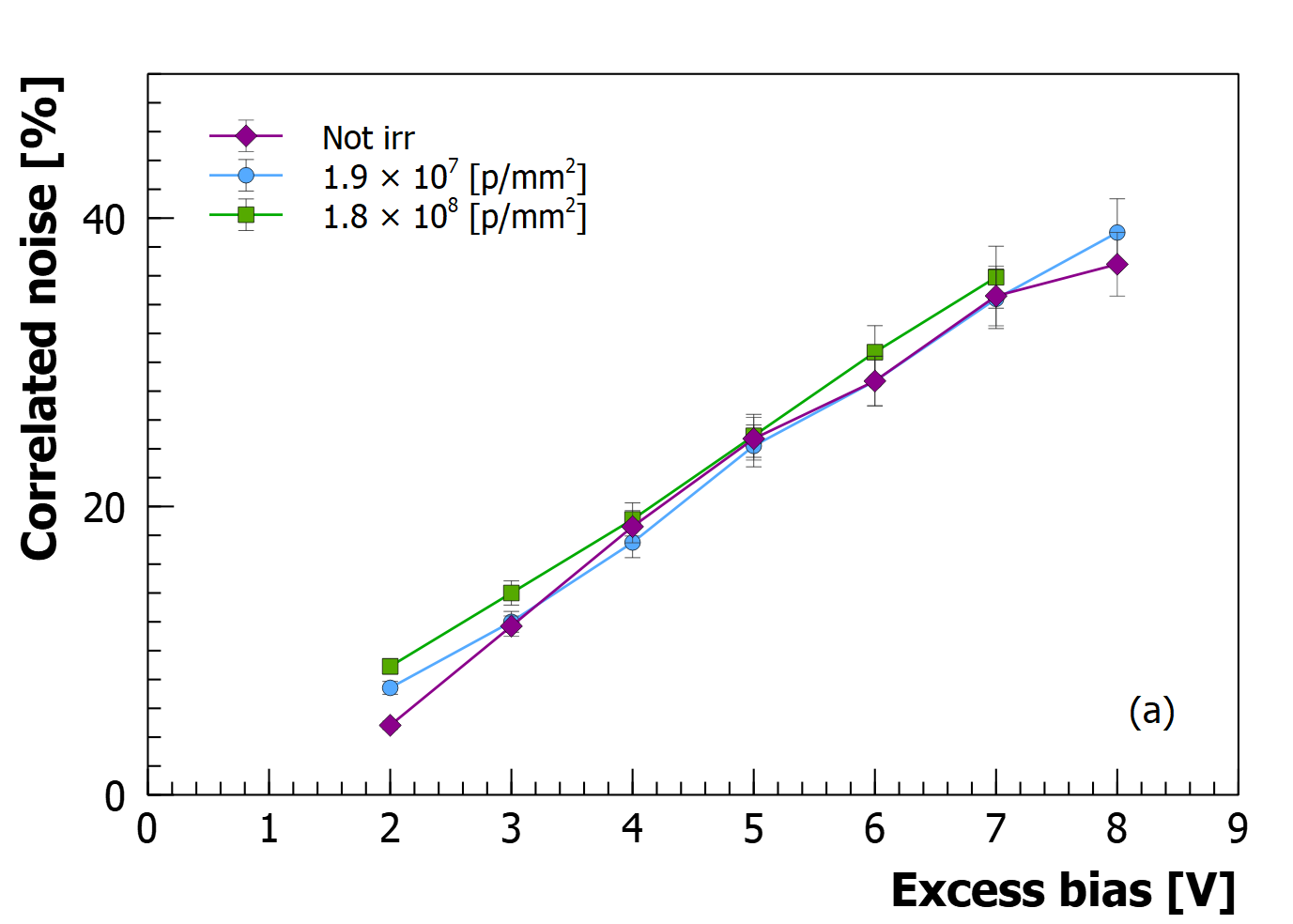}
	\end{subfigure}
	\begin{subfigure}{.45\textwidth}
		\includegraphics[width=1\textwidth]{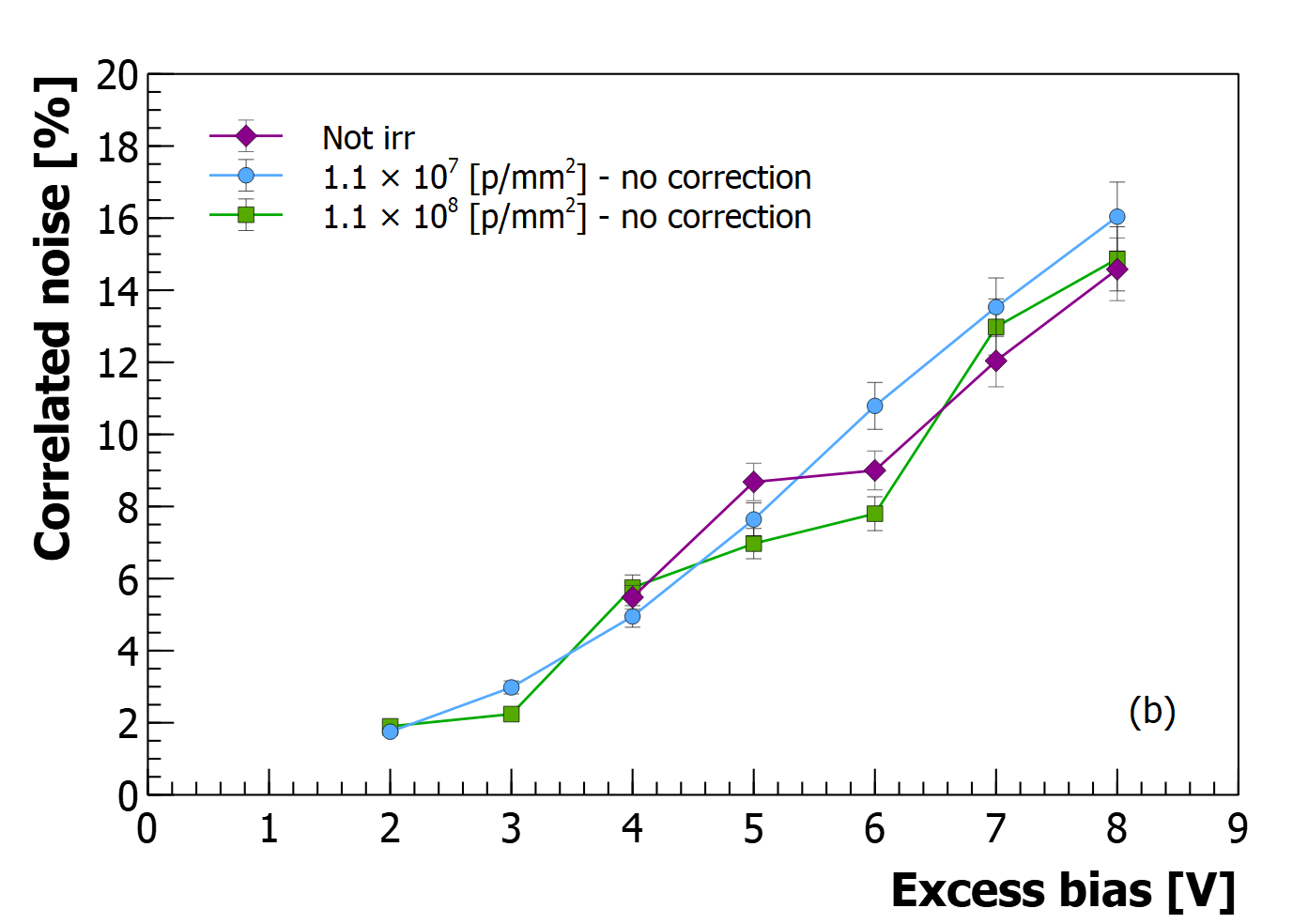}
	\end{subfigure}
	\caption{Correlated noise of the RGB-HD SiPM SiPM with 25$\mu$m cell pitch (a) and NUV-HD-RH SiPM with 15$\mu$m cell pitch (b) at two fluence levels.}
	\label{fig:correla}
\end{figure}

The correlated noise as sum of direct cross-talk, delayed cross-talk and afterpulsing probabilities plays a central role in the performance of a SiPM. They were measured with the pulse-identification method, thus we have results only at the lowest fluences. In Fig.\ref{fig:correla} the results for the RGB-HD SiPM and the NUV-HD-RH SiPM are visible, showing a quite coherent trend at all the fluence levels taken into account. It would be interesting to know if this consistency is still valid for higher fluences but in this sense a new measure approach needs to be investigated.

\begin{figure}[tb]
	\centering
	\begin{subfigure}{.45\textwidth}
		\includegraphics[width=1\textwidth]{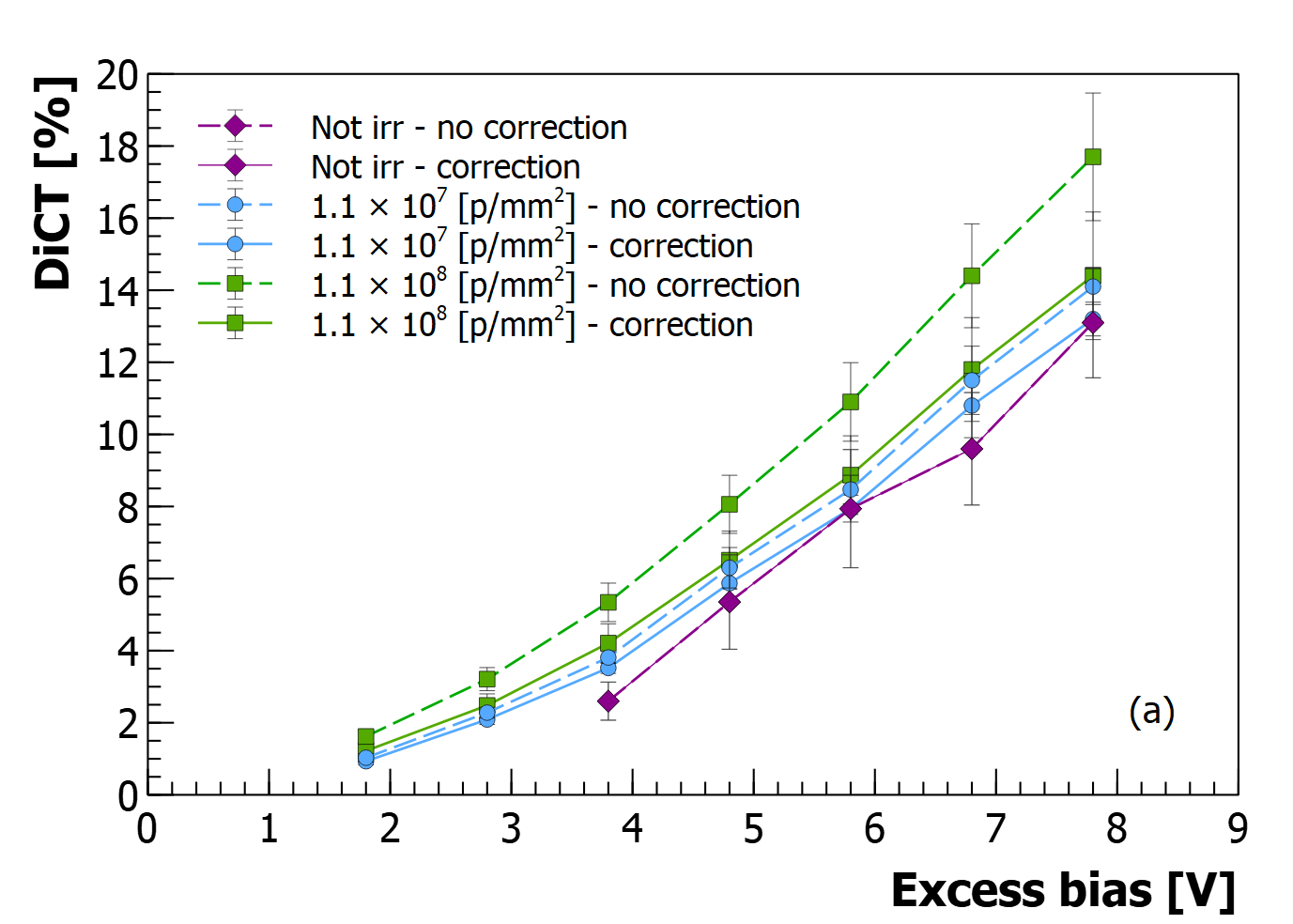}
	\end{subfigure}
	\begin{subfigure}{.45\textwidth}
		\includegraphics[width=1\textwidth]{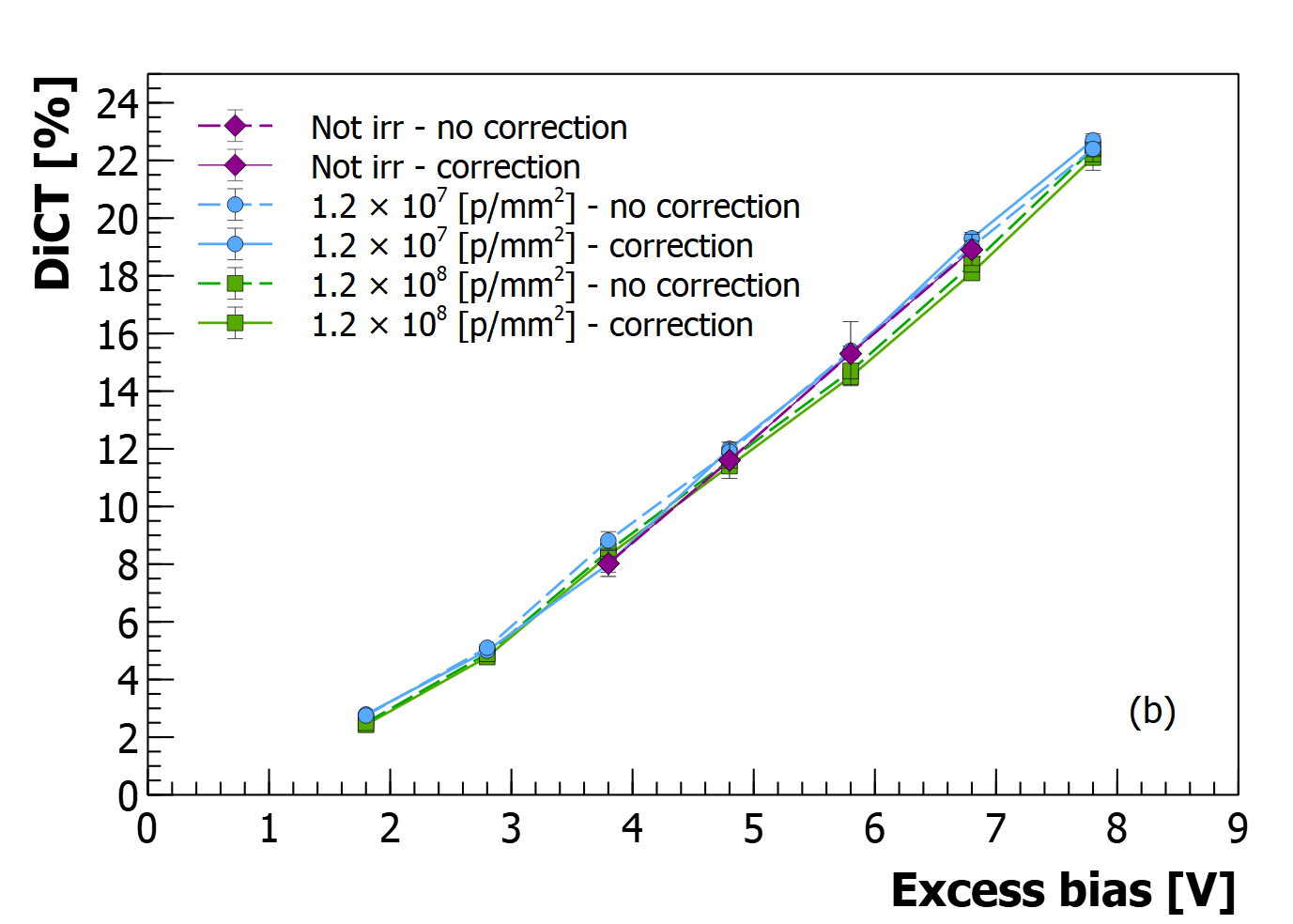}
	\end{subfigure}
	\caption{DiCT with and without the correction factor for the NUV-HD-RH SiPM with 15$\mu$m cell pitch (a) and NUV-HD SiPM with 35$\mu$m cell pitch (b).}
	\label{fig:DiCTnuvrh}
\end{figure}

The calculation of the DiCT in regimes where the DCR is very high deserves a detailed discussion. Indeed, when the total DCR (as well as the primary DCR) is very high, it is possible that two primary noise events happening very close together (i.e. closer than the minimum time interval that can be resolved by the acquisition system) are not distinguished by the pulse-identification algorithm and are considered as a unique event with va double amplitude, i.e. they are erroneously labeled as a cross-talk event. The mislabelling probability of such kind of issue increases with increasing DCR and decreases with increasing bandwidth of the acquisition system.
In our experimental setup, we commonly use an analog bandwidth of 1 GHz and the DLED algorithm \cite{GolaDLED2013} to improve the peak-finding capabilities, to reduce the effects of the pile-up of subsequent pulses. Therefore, we used a correction factor for the direct cross-talk (DiCT) calculation. Using the formula reported in \cite{CT.formula} with a minor modification. In our case, the not-corrected cross-talk was calculated as the ratio of the count rate (or equivalently the probability) of the 2 p.e. peak (i.e. events between 1.5 p.e. and 2.5 p.e.) over the count rate of the 1 p.e. peak (i.e. events between 0.5 p.e. and 1.5 p.e.), as reported in \cite{dcr}. The result is shown in the following formula for the corrected cross-talk:
\begin{equation}
    p_{CT}=1-\Bigg[1-\frac{P_{2 p.e.}}{P_{1 p.e.}}\Bigg]\times e^{DCR_{0.5}\times \tau}
\end{equation}
where $\tau$ is the minimum inter-arrival time that can be distinguished by the acquisition system and $DCR_{0.5}$ represents the total DCR, measured with a 0.5 p.e. threshold level.
Some examples are reported in Fig.\ref{fig:DiCTnuvrh}, showing the DiCT at the two lowest fluences before and after the correction factor for the NUV-HD-RH SiPM with a 15$\mu$m cell pitch and the NUV-HD-RH SiPM with a 35$\mu$m cell pitch. A significant change in the curves can be noticed due to the correction factor in the NUV-HD-RH SiPM, showing a DiCT constant with fluence at least up to $10^8$ $p/mm^2$, while no remarkable effects were identified in this sense for the other technologies tested, as visible in Fig.\ref{fig:DiCTnuvrh} for the NUV-HD SiPM.

\subsection{Activation Energy}

\begin{figure*}[tb]
	\centering
	\begin{subfigure}{.33\textwidth}
		\includegraphics[width=1\textwidth]{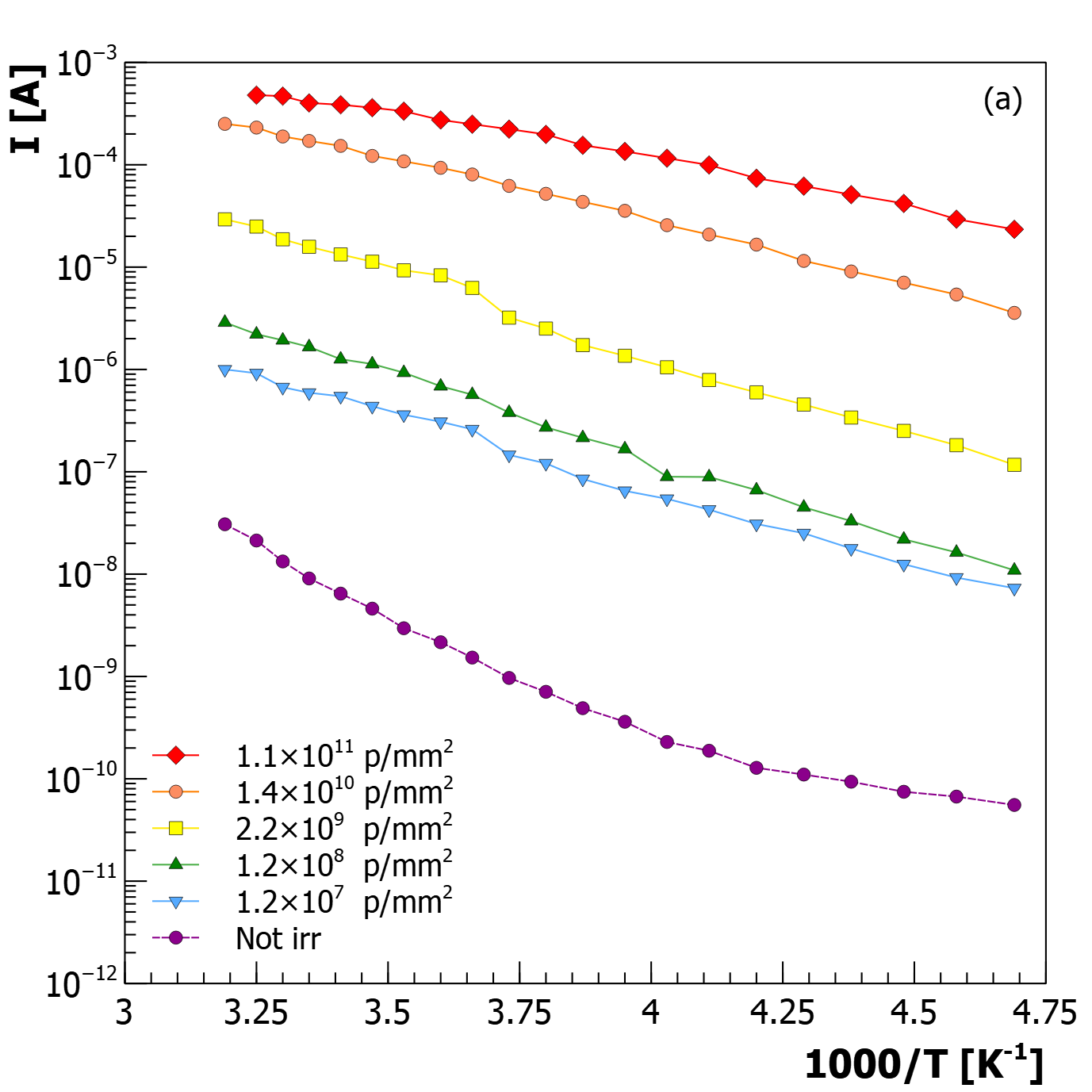}
	\end{subfigure}
	\begin{subfigure}{.33\textwidth}
		\includegraphics[width=1\textwidth]{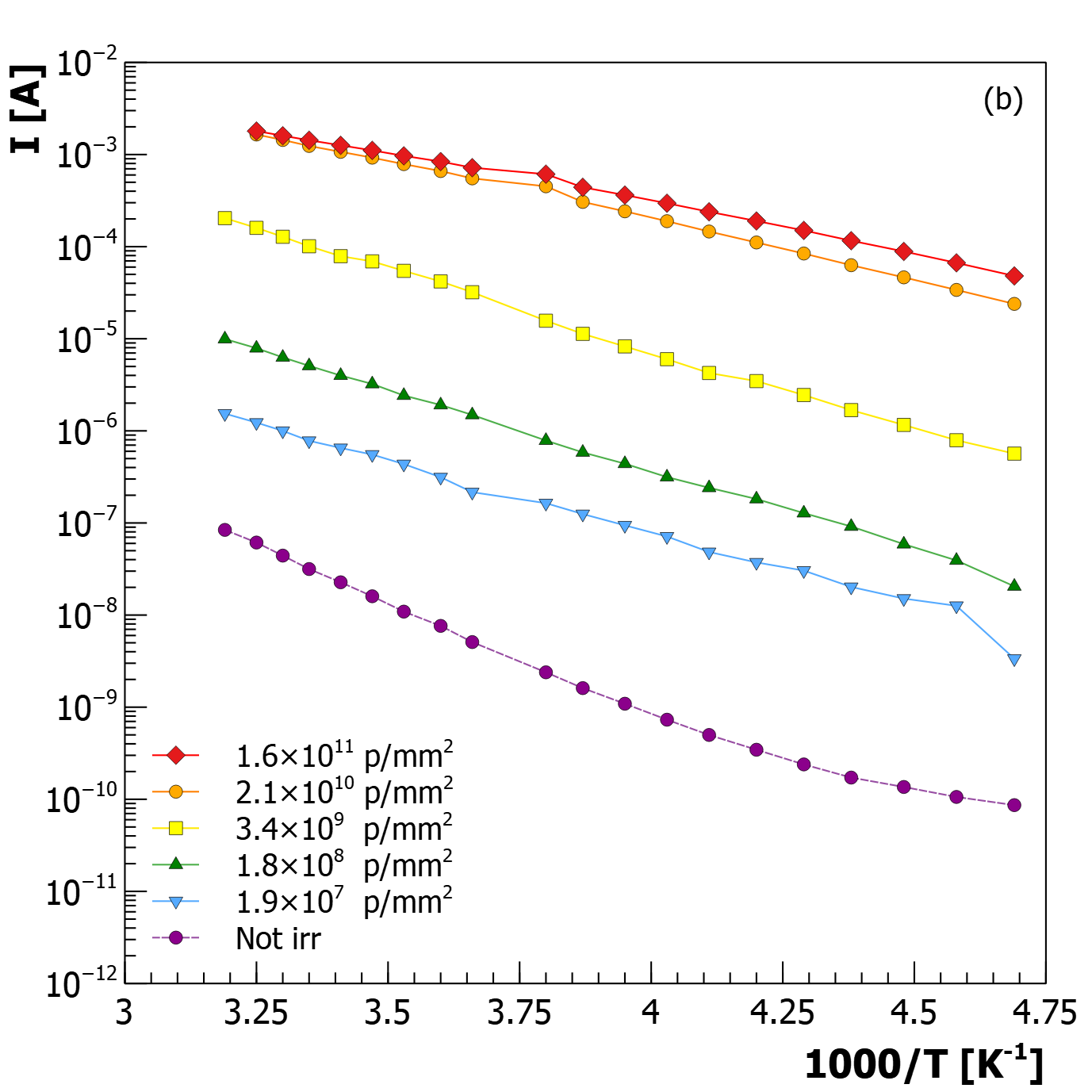}
	\end{subfigure}
	\begin{subfigure}{.33\textwidth}
		\includegraphics[width=1\textwidth]{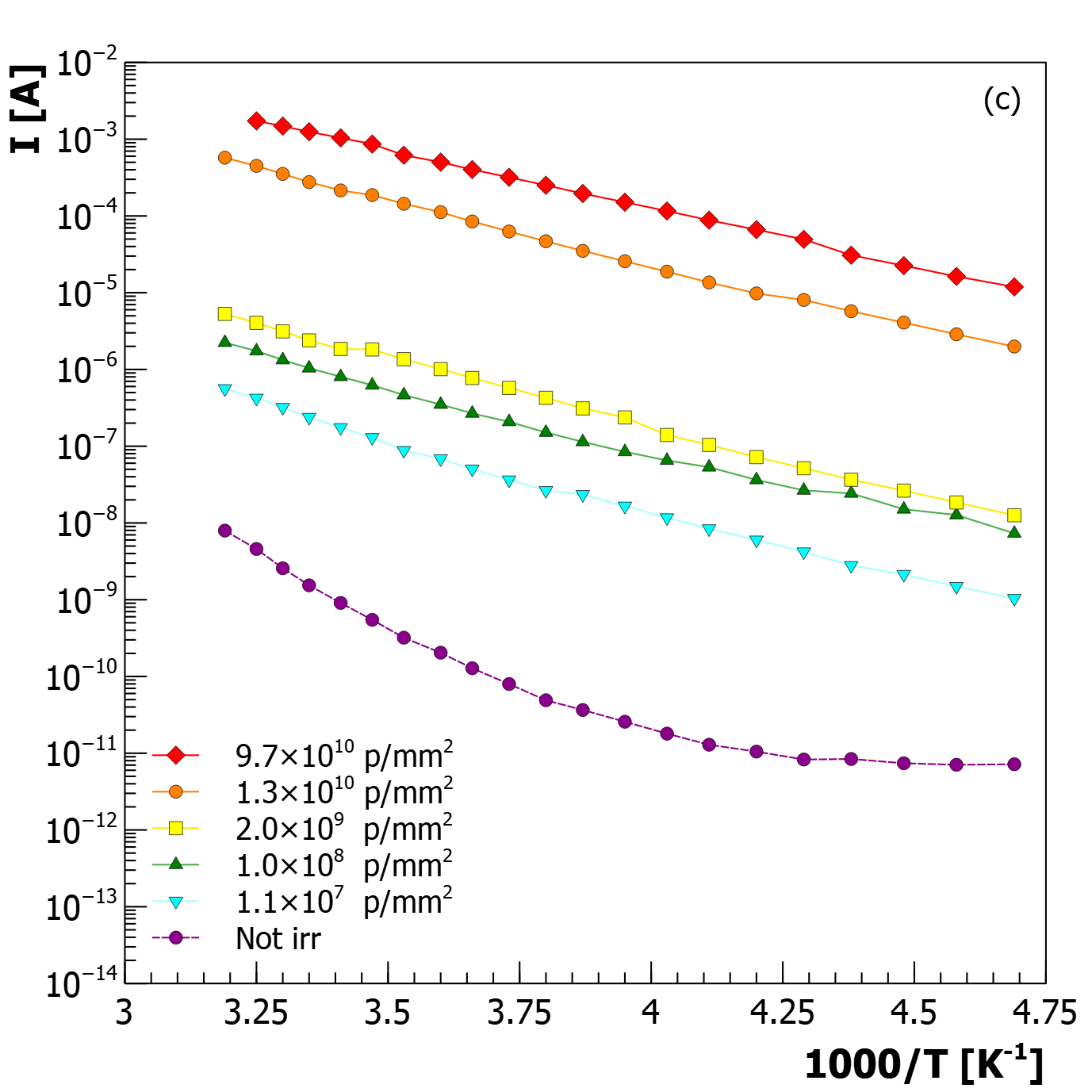}
	\end{subfigure}
	\caption{Reverse current as function of the temperature for the NUV-HD with 35$\mu$m cell pitch (a), RGB-HD with 25$\mu$m cell pitch (b) and NUV-HD-RH with 15$\mu$m cell pitch (c) technologies.}
	\label{fig:IvsT}
\end{figure*}

\begin{figure*}[tb]
	\centering
	\begin{subfigure}{.33\textwidth}
		\includegraphics[width=1\textwidth]{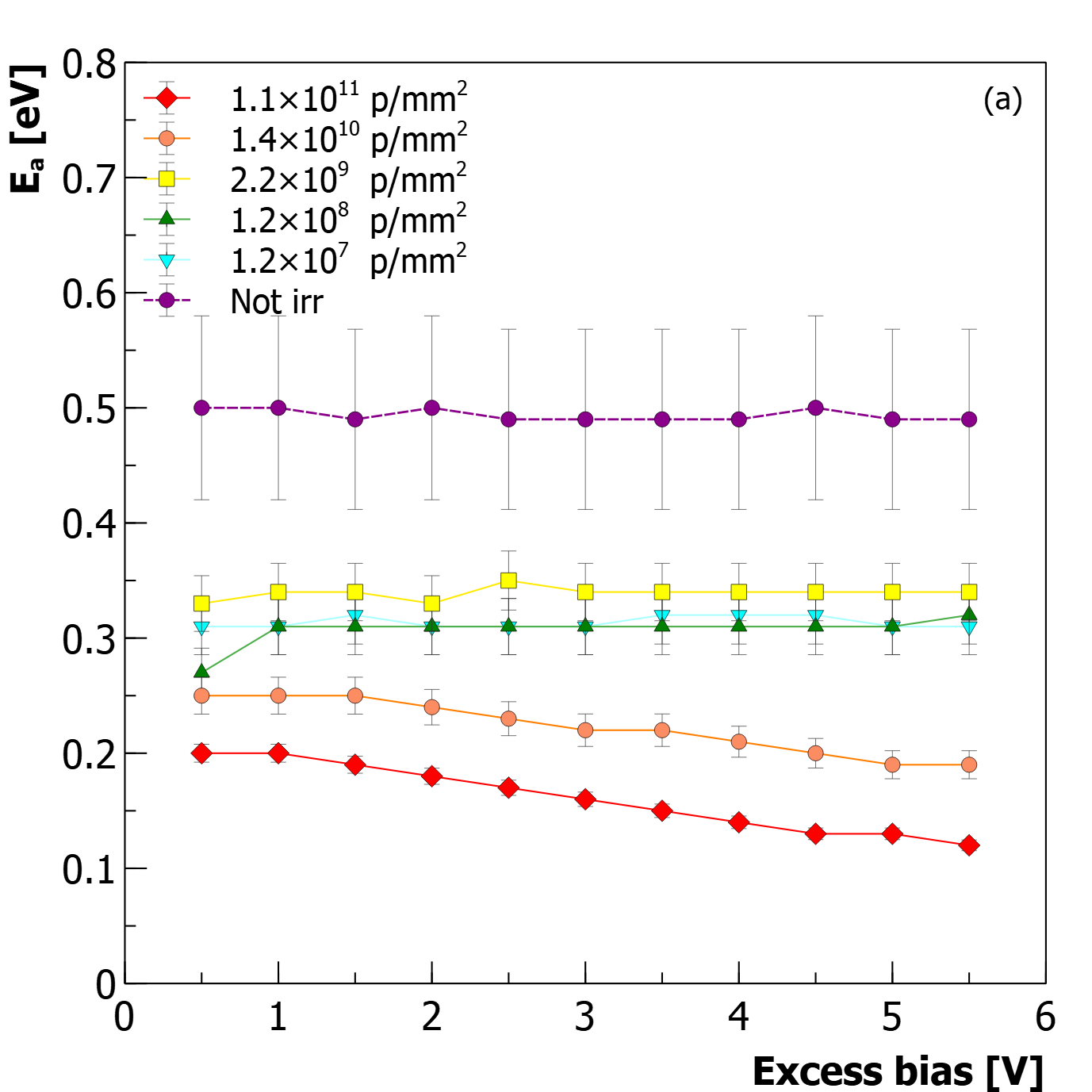}
	\end{subfigure}
	\begin{subfigure}{.33\textwidth}
		\includegraphics[width=1\textwidth]{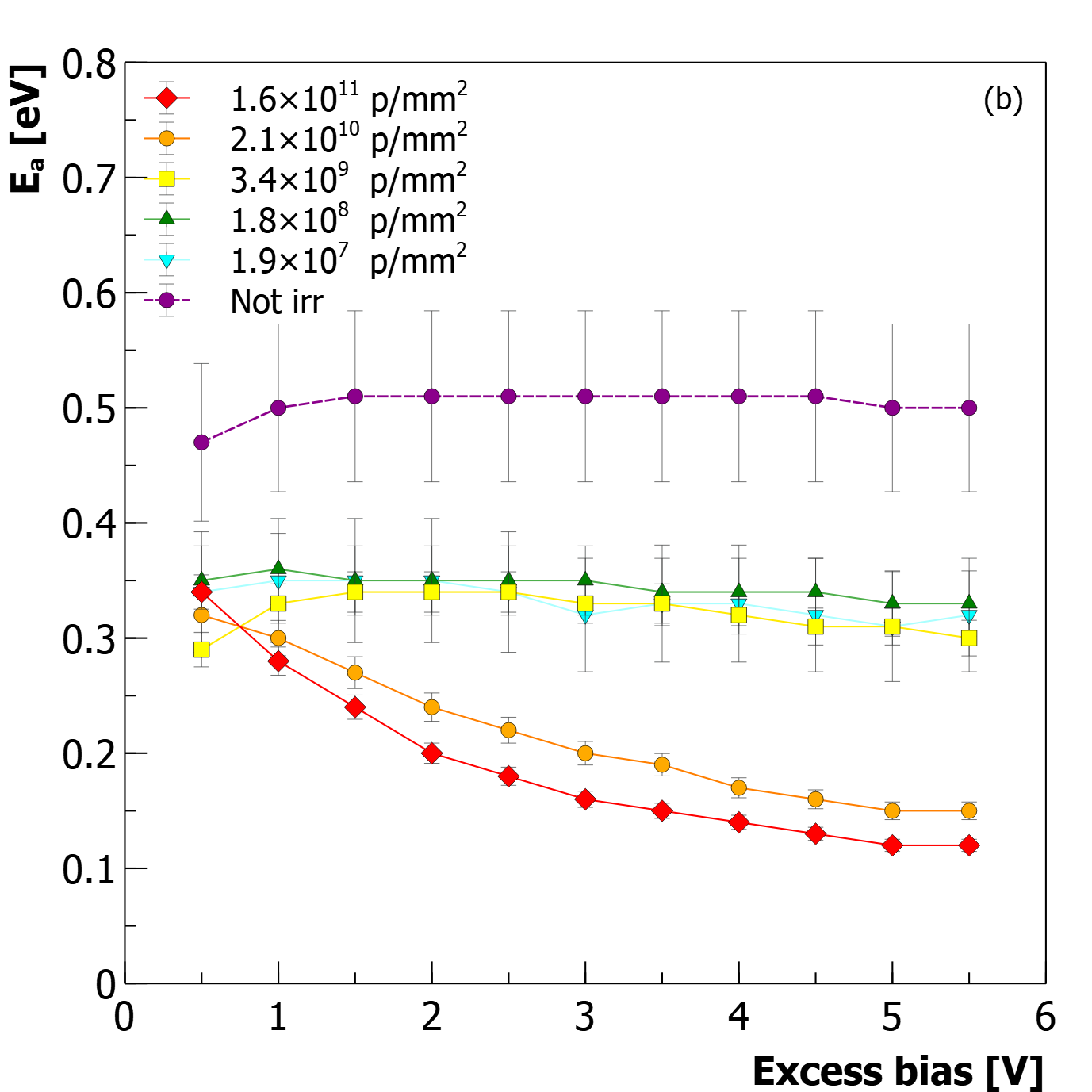}
	\end{subfigure}
	\begin{subfigure}{.33\textwidth}
		\includegraphics[width=1\textwidth]{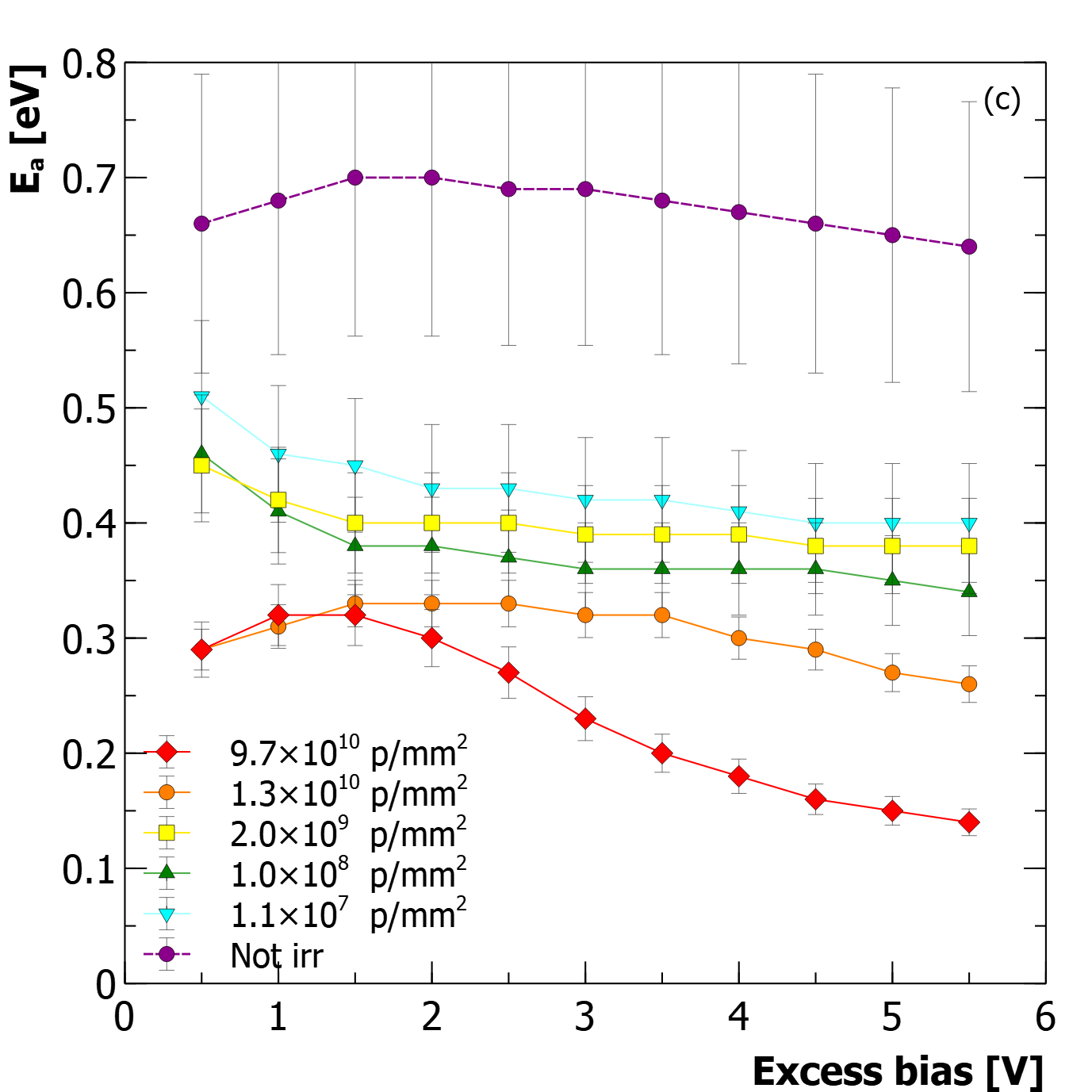}
	\end{subfigure}
	
	\caption{Activation energy (Ea) as function of the excess bias for the NUV-HD with 35$\mu$m cell pitch (a), RGB-HD with 25$\mu$m cell pitch (b) and NUV-HD-RH with 15$\mu$m cell pitch (c) technologies.}
	\label{fig:EavsOV}
\end{figure*}

We also characterized the SiPMs at different temperatures, by means of a climatic chamber. In particular we measured the current-voltage curves and we extracted the slope of the current as a function of the temperature, thus obtaining the activation energy. 


The dark current has an Arrhenius-like dependence on the temperature described by the relation:
\begin{equation}
    I(T)=I_{0} e^{-\frac{E_a}{kT}}
\end{equation}

The Activation Energy ($E_a$) can be extracted from the I(T) function. Measurement were performed in the range $-50^{\circ}\div 35^{\circ}$, as shown in Fig.\ref{fig:IvsT}.
The fit of I(T) to extract $E_a$ was performed in the range between -$15^{\circ}C$ and +$15^{\circ}C$ where a good linearity was ensured and where we are confident that thermal generation is the dominant mechanism. \\
$E_a$ is a key parameter to obtain an accurate interpretation of the microscopic effects of the damage inside the sensor. 
In Fig.\ref{fig:EavsOV} the extrapolated activation energy as a function of the excess bias is plotted for the NUV-HD with 35$\mu$m cell pitch, RGB-HD with 25$\mu$m cell pitch and NUV-HD-RH with 15$\mu$m cell pitch SiPMs. Some saturation effects of the reverse current are visible starting from fluences in the order of $10^{10} p/mm^2$. This led to a fictitious 
decrease of $E_a$. This is more noticeable in the NUV-HD SiPM with 35$\mu$m cell pitch where the beginning of the saturation is perfectly visible even at low excess biases and the decrease of the $E_a$ with the fluence becomes clear. Excluding the curves affected by saturation effects, at least two $E_a$ levels can be outlined in the plot when the excess bias is fixed, showing a clear decrease of the activation energy from the value of the not-irradiated sample (i.e. $0.5 \div 0.65$ eV) to the value of the samples irradiated at $10^7 \div 10^9$ $p/mm^2$ (i.e. about $0.35 \div 0.4$ eV). These results have to be still fully understood, but this behaviour might suggest the generation of some additional energy levels, lowering the activation energy from around mid-gap (i.e. around 0.6 eV) to a lower value, around 0.3 eV. This might be possibly due to the creation of vacancies and interstitial atoms due to the radiation effects.

\begin{figure*}[tb]
	\centering
	\begin{subfigure}{.33\textwidth}
		\includegraphics[width=1\textwidth]{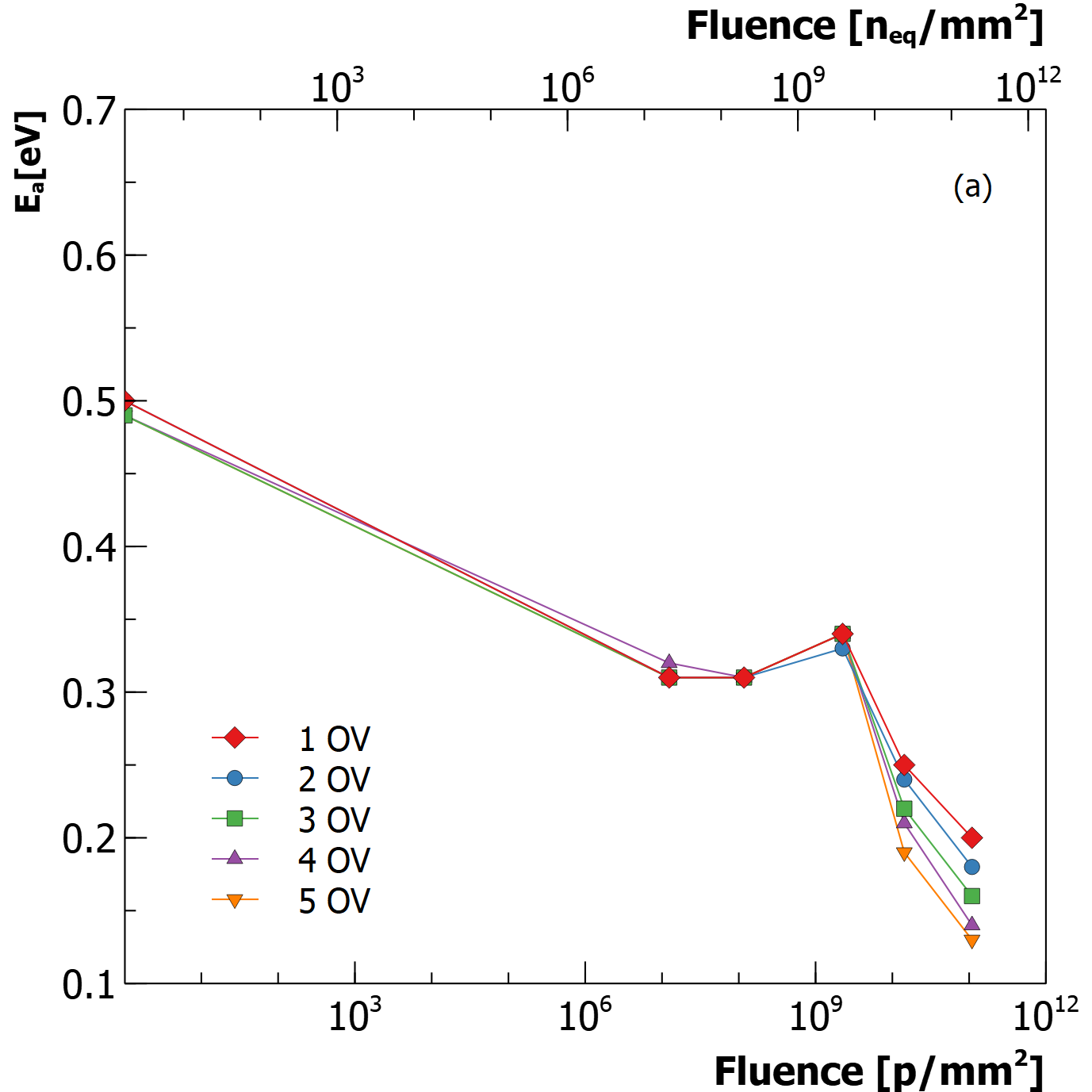}
	\end{subfigure}
	\begin{subfigure}{.33\textwidth}
		\includegraphics[width=1\textwidth]{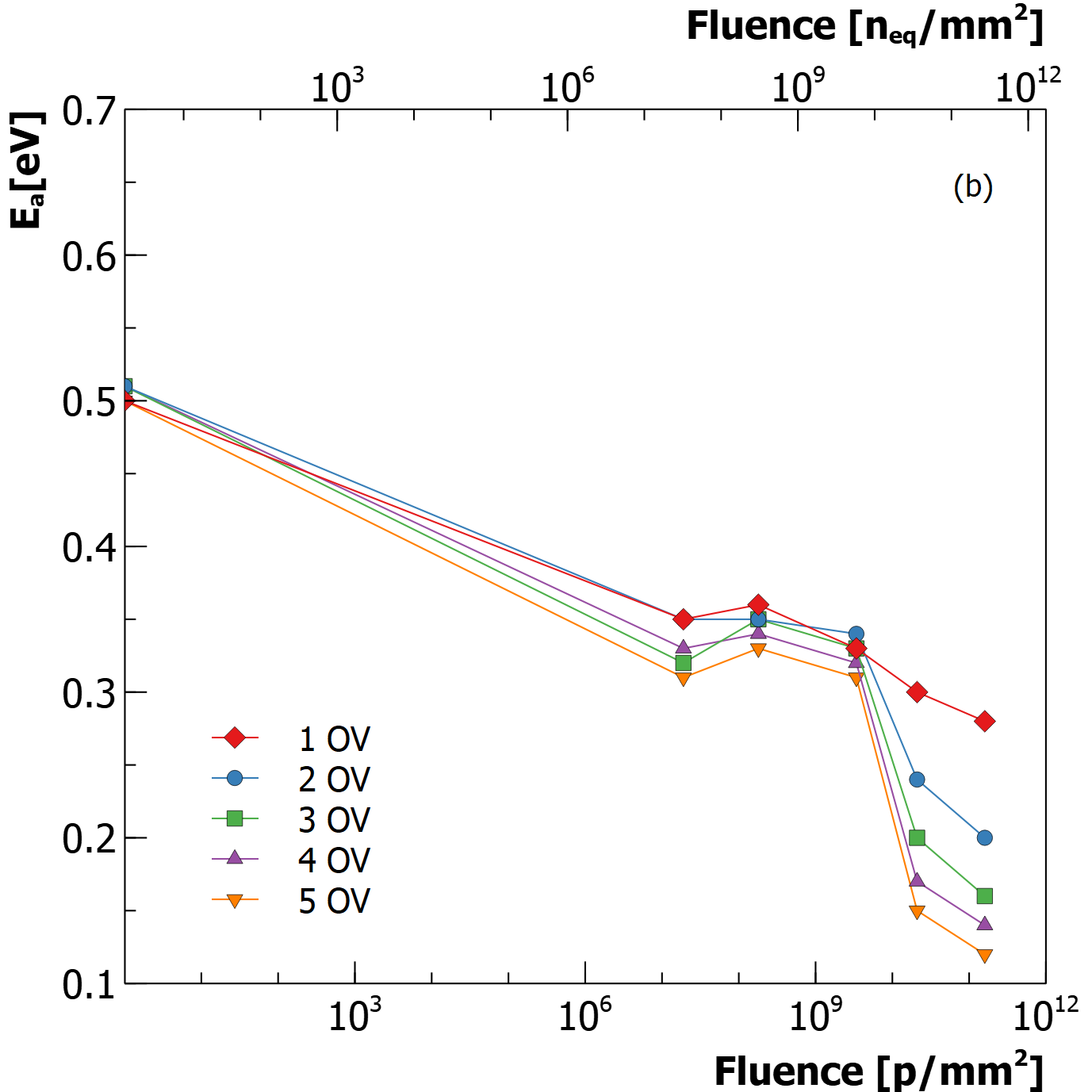}
	\end{subfigure}
	\begin{subfigure}{.33\textwidth}
		\includegraphics[width=1\textwidth]{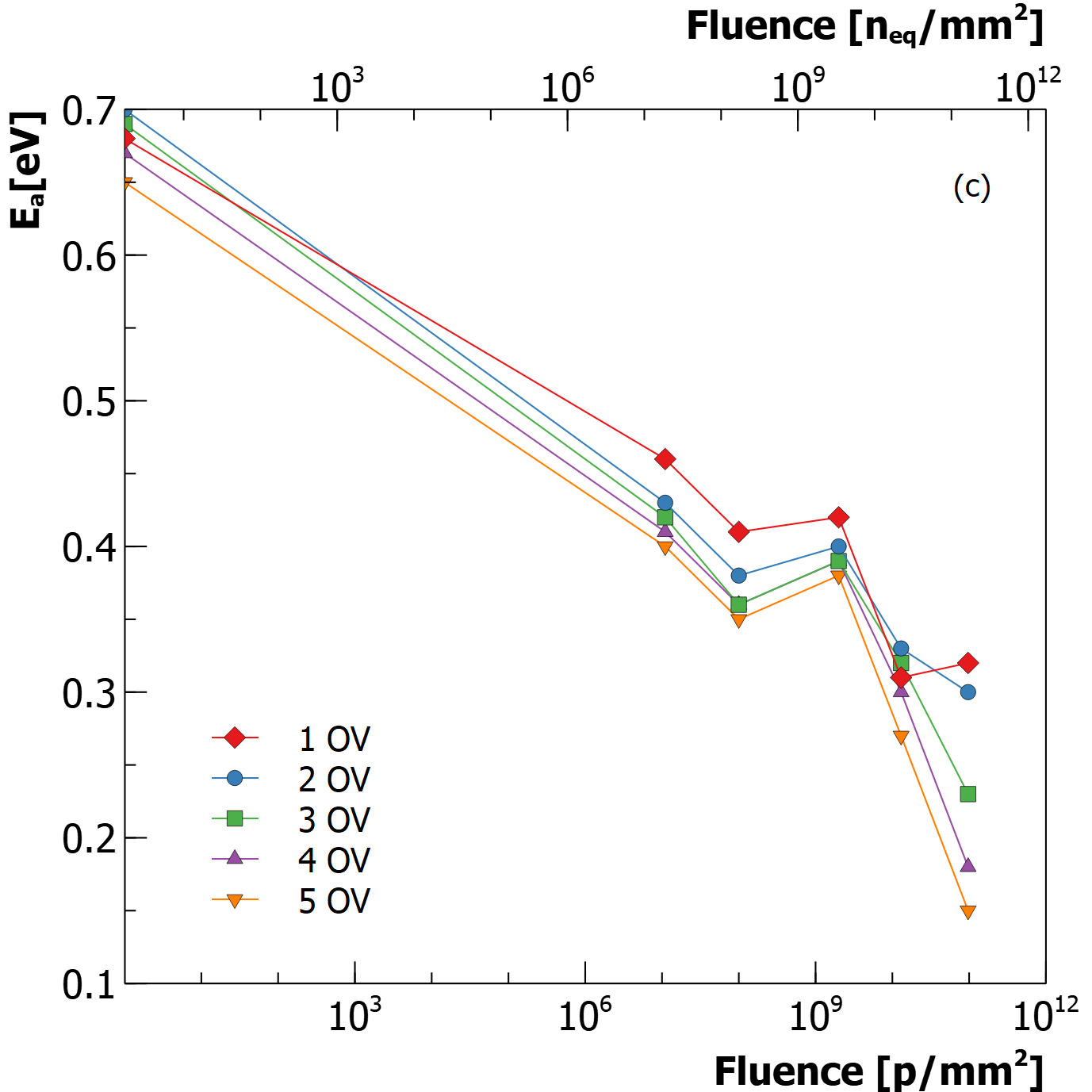}
	\end{subfigure}
	\caption{Ea as function of the $\Phi$ for the NUV-HD with 35$\mu$m cell pitch (a), RGB-HD with 25$\mu$m cell pitch (b) and NUV-HD-RH with 15$\mu$m cell pitch (c) technologies.}
	\label{fig:EavsF}
\end{figure*}

The saturation effect is also visible in Fig.\ref{fig:EavsF} where a gradual decrease of $E_a$ can be observed. The growth of the gap among the excess biases with increased fluences reveals an almost complete saturation at $10^{11}p/mm^2$.

\section{Emission microscopy}\label{sec:emmi}
Emission Microscopy (EMMI) is a technique originally developed for VLSI failure analysis. It is based on imaging of "hot carrier luminescence" (HCL), which is due to accelerated carriers suddenly losing their energy in high electric field regions. In SPADs and SiPMs the emission of secondary photons happens during the avalanche multiplication process. This can be ascribed to different mechanisms \cite{Akil1999,Gautam,lacaita}.

\begin{figure}[tb]
    \centering
    \includegraphics[width=0.47\textwidth]{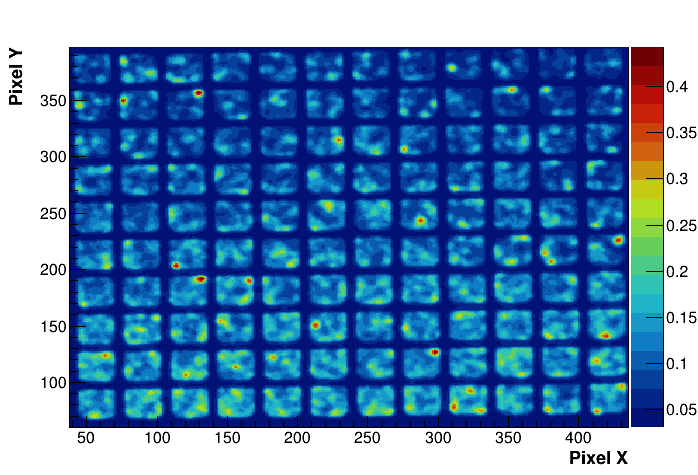}
    \caption{EMMI capture of the NUV-HD SiPM with 35$\mu$m cell pitch, fed at 2.5V excess bias with a zoom on a low number of cells.}
    \label{fig:EMMIwholeNUV}
\end{figure}

\begin{figure}[tb]
    \centering
    \includegraphics[width=0.47\textwidth]{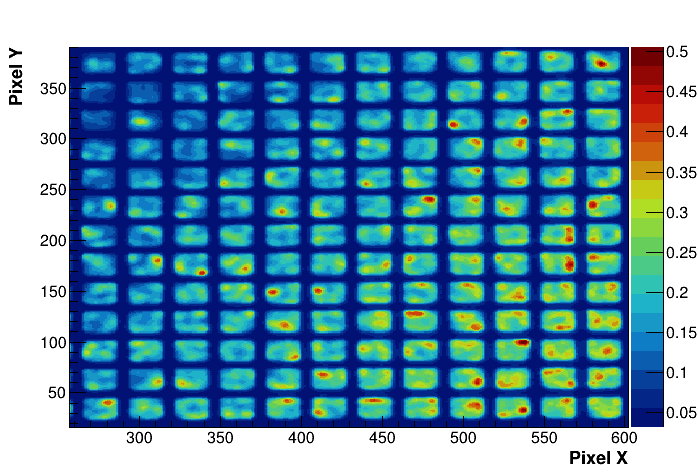}
    \caption{EMMI capture of the RGB-HD SiPM with 30$\mu$m cell pitch, fed at 3.5V excess bias with a zoom on a low number of cells.}
    \label{fig:EMMIwholeRGB}
\end{figure}

In our case, we tested some of the most irradiated devices and we focused on their enhanced-light regions which should correspond to the regions where most likely and most often a ``noise avalanche pulse'' is generated. Regions with an high light emission can be found within a single cell of the SiPM, distributed in the whole structure, as visible in Fig.\ref{fig:EMMIwholeNUV} and Fig.\ref{fig:EMMIwholeRGB}, that show the EMMI images of the NUV-HD SiPM with 35$\mu$m cell pitch and the RGB-HD SiPM with 25$\mu$m cell pitch, respectively, irradiated at $10^{10}$ $p/mm^2$. The detected light was directly used for the estimation of the DCR in other works \cite{engelmann}.

In this work, in order to understand the main failure mechanisms of the SiPMs and to be able to possibly improve their radiation hardness, we focused on finding a preferred spatial localization of the enhanced light emission regions into the single cells of the SiPM. To do that, the EMMI image was considered as a matrix of intensity points $I_{i,j}$. An ellipse was created in the center of each single cell of the SiPM to divide it into two main regions: the border and the center. Then, an intensity threshold $I_{th}$ was considered and the high intensity points in the internal and external region $I_{i,j} \geq I_{th}$ were counted. 

\begin{equation}
  N_{points} = \sum{n_{points}} \:\:\:\:\:\:with \: I_{i,j} \geq I_{th}
\end{equation}

This was done for several intensity thresholds and different ellipse sizes. The main purpose of this approach was to make a preliminary statistic of the spatial localization of the enhanced light regions for different light intensities, without taking into account the dimension of the hotspots.

\begin{figure}[tb]
    \centering
    \includegraphics[width=0.45\textwidth]{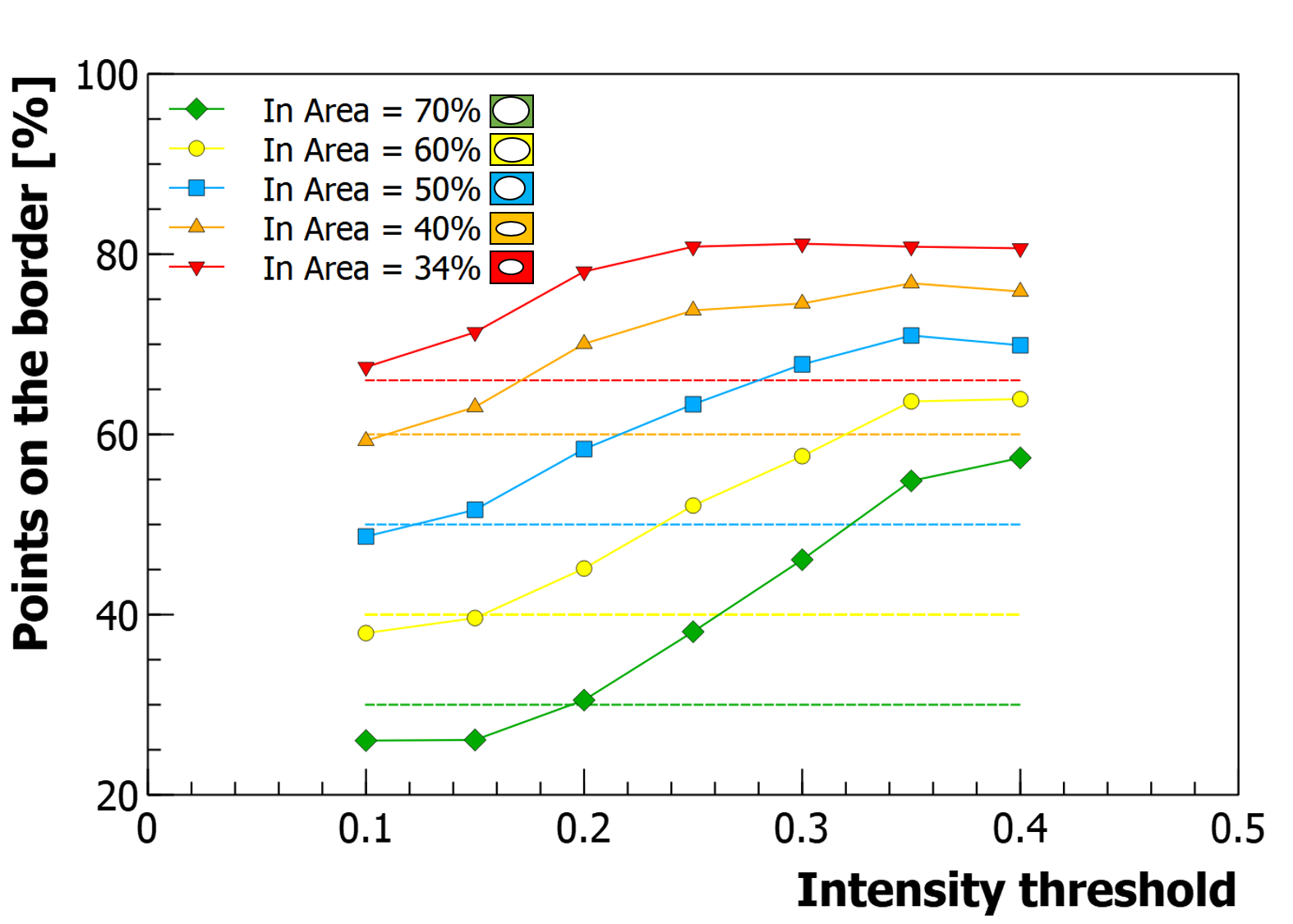}
    \caption{Plot of the points on the borders of the cell of the NUV-HD SiPM with 35$\mu$m cell pitch at 2.5V of excess bias. The dashed lines represent the values it would take if the points were equally-distributed inside the cell.}
    \label{fig:EMMIborders}
\end{figure}

\begin{figure}[tb]
    \centering
    \includegraphics[width=0.45\textwidth]{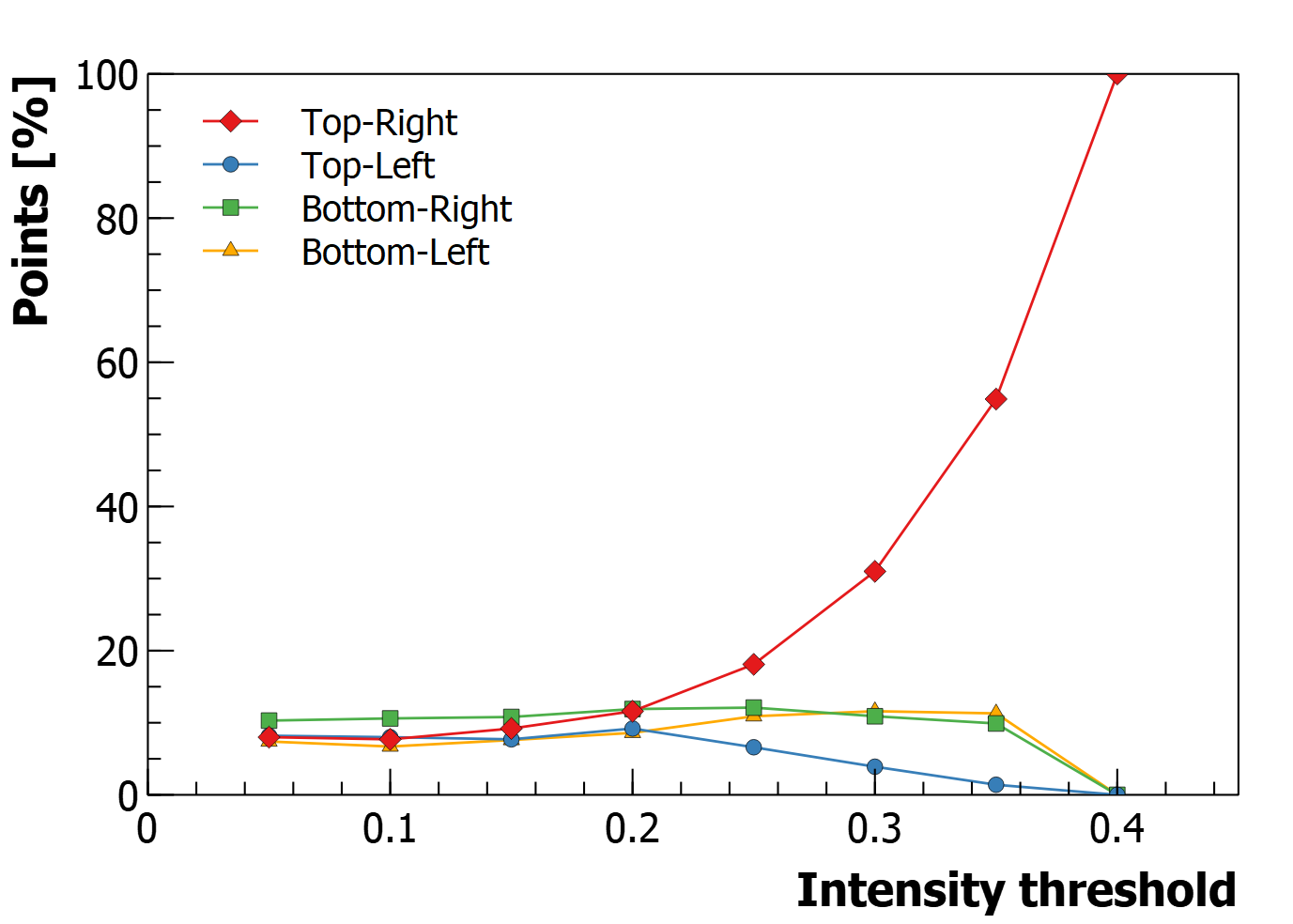}
    \caption{Percentage of the points distributed on the four angles of the cell of the NUV-HD SiPM with 35$\mu$m cell pitch at 2.5V of excess bias as function of the intensity threshold.}
    \label{fig:EMMIangles}
\end{figure}

The plot in Fig.\ref{fig:EMMIborders} shows the emission intensity of the points outside the internal ellipse (i.e. on the border of the cell) as a function of the light intensity threshold. To make a comparison, the dashed lines represent the ideal situation in case of equally distributed spots all over the cell area. We can see that the high-intensity emission points seem to have a preferred spatial localisation on the borders of the cells. A rather uniform distribution of the light regions over the entire cell was observed for low $I_{th}$ which probably is not highly relevant, as it mostly represents the background light. Furthermore, a preliminary statistic on the localization of enhanced-light points on the four corners of the cell was performed as visible in Fig.\ref{fig:EMMIangles}. Here it appears to have a preference for the top-right angle at high intensity thresholds for the NUV-HD SiPM with 35$\mu$m cell pitch. However this result is still under evaluation and yet to be completely understood. 

Future improvements of this method could hopefully lead to a more accurate definition of the spatial localization of the points and to a relation between the enhanced light points and the damage effects of the radiation in the SiPMs.

\section{Conclusions}
We performed irradiation tests on several FBK silicon photomultipliers, at the INFN-LNS in Catania (Italy) with protons at about 62 MeV and at different fluences, up to approximately $8\times10^{11} p/mm^2$ (i.e. $8\times10^{13} p/cm^2$). All irradiated SiPMs are still working up to the maximum fluence, with minor modification in most of their functional parameters, except for the primary noise and the detection efficiency.

Starting from $10^9 p/mm^2$ a significant increase of both leakage and dark current was observed. Due to the difficulty of the pulse counting at high noise levels, two methods were used for the estimation of the DCR: a first approach is based on the inter-arrival time of the pulses and a second one is based on the variation of the measured reverse dark current. For the validation of the second approach, we estimated the correlated noise by means of average signal when the SiPM was illuminated with pulsed light, extracting the gain and the excess charge factor, at the difference fluences. As a result, some saturation effects were found in DCR at high fluences, indicating a high cell occupancy, i.e. the noise reaching the limits of the SiPM dynamic range. This is less relevant in the SiPMs with smaller cell pitches, thus higher density of cells. Overall, no clear trends were found when comparing the DCR values in all the technologies. This suggests that all the devices reach a similar damage level. A key feature to improve their radiation hardness could be the study of their response to the temperature or their annealing with temperature and time, to obtain some more information about the evolution of the damage in each technology. 

The breakdown voltage is generally not straightforward to estimate in SiPMs with high leakage and dark current. This was done using several methods. The results showed some discrepancies, but overall we did not see any clear trend of the  breakdown voltage with the irradiation fluence on all samples up to $10^{11}p/mm^2$.
Measurement of the reverse current over temperature, in the range $(-60\div40)^{\circ}C$, showed an estimated activation energy clearly decreasing as the irradiation fluence increases. 

Furthermore, we also performed an emission microscopy analysis of the irradiated SiPMs, particularly focusing on a possible preferred localization of the damages and the noise avalanche generation centers. The obtained images showed an increased activity and emission in irradiated samples over the entire area of the micro-cells but with the most intense spots mostly located on the borders. An accurate analysis on the hotspot localization, with respect to the micro-cell layout is currently ongoing. 
 
\section*{Acknowledgement}
The authors thank A. Kratz and C. Karagiannis for the technical support provided during the preparation of the experiment and the technical staff of the INFN-LNS for the valuable work done for delivering the beam.






\bibliographystyle{model1-num-names}
\bibliography{sample.bib}







\end{document}